\documentclass[11pt]{article}
\usepackage{jcappub}
\usepackage{array, natbib, amsfonts}
\usepackage{amssymb, amsmath}
\usepackage{epsfig}
\usepackage{graphicx,color}
\pdfoutput=1

\newcommand{\hMpc}{{$h^{-1}$Mpc}}
\newcommand{\hMpcinv}{{$h\,{\rm Mpc}^{-1}$}}

\newcommand{\avgplus}{{\langle \gamma_+ \rangle}}
\newcommand{\avgplust}{{\langle {\gamma}_+ \rangle^{\rm trac}_{r_p}}}

\newcommand{\Pl}{P_{\rm lin}}
\newcommand{\Plsq}{{P_{\rm lin}^2}}

\newcommand{\Leg}{{\mathcal{P}}}

\newcommand{\plotdira}{Figures/}

\newcommand{\mnras}{{\rm MNRAS}}
\newcommand{\apj}{{\rm ApJ}}

\newcommand{\jcap}{{\rm JCAP}}
\newcommand{\aap}{{\rm Astron.Astrophys.}}

\newcommand{\prd}{{\rm Phys.Rev.D}}
\newcommand{\apjl}{{\rm ApJL}}
\newcommand{\aj}{{\rm AJ}}
\newcommand{\physrep}{{\rm Phys.Rep.}}
\title{Tidal alignment of galaxies}
\author[a]{Jonathan Blazek,}
\author[b,c]{Zvonimir Vlah,}
\author[d]{and Uro\v{s} Seljak}

\affiliation[a]{Center for Cosmology and AstroParticle Physics, Department of Physics,\\The Ohio State University, Columbus, OH, USA}
\affiliation[b]{Stanford Institute for Theoretical Physics and Department of Physics, Stanford University, Stanford, CA, USA}
\affiliation[c]{Kavli Institute for Particle Astrophysics and Cosmology, SLAC and Stanford University, Menlo Park, CA, USA}
\affiliation[d]{Departments of Physics and Astronomy and Lawrence Berkeley National Laboratory,\\University of California, Berkeley, CA, USA}
\emailAdd{blazek@berkeley.edu}
\abstract{We develop an analytic model for galaxy intrinsic alignments (IA) based on the theory of tidal alignment. We calculate all relevant nonlinear corrections at one-loop order, including effects from nonlinear density evolution, galaxy biasing, and source density weighting. Contributions from density weighting are found to be particularly important and lead to bias dependence of the IA amplitude, even on large scales. This effect may be responsible for much of the luminosity dependence in IA observations. The increase in IA amplitude for more highly biased galaxies reflects their locations in regions with large tidal fields. We also consider the impact of smoothing the tidal field on halo scales. We compare the performance of this consistent nonlinear model in describing the observed alignment of luminous red galaxies with the linear model as well as the frequently used ``nonlinear alignment model,'' finding a significant improvement on small and intermediate scales. We also show that the cross-correlation between density and IA (the ``GI'' term) can be effectively separated into source alignment and source clustering, and we accurately model the observed alignment down to the one-halo regime using the tidal field from the fully nonlinear halo-matter cross correlation. Inside the one-halo regime, the average alignment of galaxies with density tracers no longer follows the tidal alignment prediction, likely reflecting nonlinear processes that must be considered when modeling IA on these scales. Finally, we discuss tidal alignment in the context of cosmic shear measurements.}

\keywords{}
\arxivnumber{}

\begin{document}
\maketitle
\flushbottom

\section{Introduction}

Coherent, large-scale correlations of the intrinsic shapes and orientations of galaxies are a potentially significant source of systematic error in weak gravitational lensing studies. Contamination from these galaxy ``intrinsic alignments'' (IA) can bias or degrade lensing science results. Since early work establishing its potential effects \cite{croft00,heavens00,catelan01,crittenden01}, IA has been examined through observations \cite{hirata07,joachimi11,blazek12,singh14}, analytic modeling \cite{catelan01, hirata04,blazek11}, and simulations \cite{schneider12, tenneti14a} - see \cite{troxel14rev} for a recent review. As our understanding has improved, IA has also emerged as a potential probe of large-scale structure as well as halo and galaxy formation and evolution \cite{schmidt12,chisari13}.

Despite recent progress, further understanding of IA is needed. The increased statistical precision of current and next-generation lensing surveys such as DES\footnote{Dark Energy Survey, https://www.darkenergysurvey.org}, KiDS\footnote{Kilo Degree Survey, http://kids.strw.leidenuniv.nl/}, HSC\footnote{Hyper Suprime-Cam, http://www.subarutelescope.org/Projects/HSC/}, Euclid\footnote{http://sci.esa.int/science-e/www/area/index.cfm?fareaid=102}, LSST\footnote{Large Synoptic Survey Telescope, http://www.lsst.org}, and WFIRST\footnote{Wide-Field Infrared Survey Telescope, http://wfirst.gsfc.nasa.gov/} will require unprecedented control of systematic uncertainties. It is likely that a combination of approaches will be employed for effective IA mitigation. We focus in this work on analytic modeling of galaxy alignments, usually relating IA to the gravitational tidal field, which is determined by the surrounding large-scale structure \cite{catelan01,hirata04}. These analytic approaches are naturally perturbative and are analogous to galaxy bias, which relates the galaxy and dark matter density fields. On sufficiently large scales, where perturbative treatment of the tidal field converges, these models should produce a simple and accurate description of IA. On smaller scales, however, a perturbative description will necessarily break down, and a qualitatively nonlinear treatment is required. To model IA on these scales, halo model prescriptions have been developed to specify the distribution of shapes and orientations for central and satellite galaxies (e.g. \cite{schneider10,joachimi13b}). These halo models can also be combined with effective tidal field models on larger scales.

Given the current challenges and uncertainties in modeling IA, especially at small scales and for the faint sources that dominate the lensing signal, it is advisable to also consider model-independent methods to mitigate IA contamination. However, more general approaches to IA mitigation tend to come at a significant cost of statical signal (e.g.\ \cite{joachimi08}). For IA marginalization, it is important to reduce the number of ``nuisance'' parameters as well as accounting for the covariance of these parameters for multiple cosmological probes. Well-understood and accurate analytic modeling allows for efficient parameterization of IA for use in large joint analyses.

First proposed in \cite{catelan01}, the tidal alignment model posits that halo and galaxy shapes (or the correlated components) are aligned with the local tidal field. This type of alignment is most likely to arise in elliptical galaxies for which angular momentum does not play a significant role in determining shape and orientation. Indeed, tidal alignment has been shown to agree with shape correlations on large scales for luminous red galaxies (LRGs), which are highly biased, elliptical, and pressure-supported \cite{joachimi11,blazek11,singh14}. For spiral galaxies, the acquisition of angular momentum, for instance through ``tidal torquing,'' generally leads to quadratic dependence on the tidal field (e.g., \cite{lee00,hirata04}). While tidal alignment may accurately capture the relevant astrophysics for certain types of galaxies, it can also be motivated through symmetry arguments. Since the tidal alignment contribution is the lowest-order function of the gravitational potential with the necessary symmetry, it is expected to dominate IA correlations on sufficiently large scales. Due to its simplicity and its relative success at describing the IA signal of highly biased, red galaxies (the only sample with well-measured shape alignments), the tidal alignment model remains an important tool.

Because of the linear relationship assumed between galaxy shape and the tidal field, tidal alignment has historically been referred to as the ``linear alignment'' model. However, it was quickly recognized that nonlinear effects are likely important on smaller scales. Attempts to include nonlinear dark matter clustering by using a nonlinear matter power spectrum yielded the ``non-linear alignment'' (NLA) model, which provides qualitative improvement in matching LRG alignment on intermediate scales \cite{hirata04,bridle07}. The NLA model has been the fiducial functional form for several weak lensing projections and analyses (e.g.\ \cite{joachimi11,heymans13}). However, once effects beyond linear order are included, referring to the ``(non)linear alignment model'' becomes ambiguous. Instead, we use the term ``tidal alignment'' to refer to any model in which the relationship between intrinsic galaxy shapes and the tidal field is linear. Quadratic alignment (e.g.\ tidal torquing) models involve a quadratic dependence on the tidal field.

Within the context of tidal alignment, three effects can produce nonlinearities in the intrinsic shape correlations: (1) nonlinear evolution of the dark matter density field, leading to nonlinear evolution in the tidal field; (2) observing the IA field at the positions of shape tracers, leading to a weighing by the tracer density; (3) a nonlinear bias relationship between the galaxy and dark matter density fields. The NLA model includes the nonlinear evolution of the dark matter density but but does not consider other nonlinear contributions. While the NLA approach may improve the model fit to data, it is not fully consistent and omits important astrophysical effects. Some work (e.g.\ \cite{kirk12}) advocates explicitly ignoring the nonlinear evolution of the density field when calculating the intrinsic ellipticity field, assuming that alignment is set at early times when the density field is closer to linear. As discussed below, this physical picture may be relevant, and the redshift at which IA is set is an important element in any model. In this work we wish to treat all nonlinear contributions on equal footing, providing a more complete description of the tidal alignment scenario. With such a framework in place, we can more convincingly examine how well tidal alignment captures the physics of IA and on what scales a nonlinear relationship between IA and the tidal field must be considered. Having a more complete treatment of possible nonlinear effects will also allow the extension of the tidal alignment model to the types of lensing sources that are actually relevant for current and upcoming studies. We now have several measurements of IA for LRGs, but these are not typical lensing sources. Thus, even if the {\it ad-hoc} NLA model is satisfactory for LRGs, it may not be for the fainter sources that will dominate the lensing signal.

We employ two distinct approaches to examine nonlinear tidal alignment. First, we use a standard perturbation theory (SPT) expansion of the density field to calculate the one-loop non-linear contributions to the tidal alignment model. Within this expansion, we examine the effect of smoothing the tidal field, an element of IA modeling which remains an important open question (e.g.\ \cite{blazek11,chisari13}). As a second approach, we show that source clustering can be effectively separated from tidal alignment itself, allowing shape correlations to be described as a product of the two effects. We then employ non-perturbative modeling to describe the tidal field around a dark matter halo, allowing a description of IA down to significantly smaller scales. In the one-halo regime, we find that recent IA measurements indicate relatively constant galaxy alignment as a function of separation, indicating processes beyond tidal alignment.

This paper is organized as follows. Section~\ref{sec:shape_formalism} outlines the relevant formalism for measuring shape correlations. Section~\ref{sec:tidal_alignment} develops the tidal alignment model, consistently treating all nonlinear contributions using SPT, and describes the impact of smoothing the tidal field. We compare the model predictions with IA measurements. Section \ref{sec:separation} explores a non-perturbative description of tidal alignment, separating the observed intrinsic shape correlations into source alignment and clustering terms. We also discuss the impact of nonlinear tidal alignment contributions in the context of cosmic shear. A discussion of the main results is in Section~\ref{sec:discussion}. The details of many of the relevant calculations are presented in an appendix. We assume a flat $\Lambda$CDM cosmology with $\Omega_m=0.279$, $\Omega_b=0.046$, $\sigma_8=0.808$, and $h=0.701$. 
\section{Formalism of intrinsic shape correlations}
\label{sec:shape_formalism}
Observed shapes can be expressed in terms of an ellipticity, $e_0$ and position angle, $\phi$, measured with respect to a chosen direction. The shape can also be decomposed into components $(e_+,e_\times)$. In terms of the position angle and minor-to-major axis ratio, $b/a$:
\begin{align}
\label{shape_decomp}
\left[\begin{array}{c}
e_+ \\
e_{\times} \end{array}
\right]
=
\left(\frac{1-(b/a)^2}{1+(b/a)^2}\right)
\left[\begin{array}{c}
\cos(2\phi) \\
\sin(2\phi)\end{array}
\right]
\equiv
e_0
\left[\begin{array}{c}
\cos(2\phi) \\
\sin(2\phi)\end{array}
\right].
\end{align}
Note that this decomposition obeys the necessary spin-2 symmetry (invariance under rotation by $n\pi$).

In the weak lensing regime, an ensemble of observed galaxy shapes provides an estimate of the lensing shear ($\gamma$). The two quantities are related through the shear responsivity $\mathcal{R}$, which captures the average response of measured ellipticity to a small shear \cite{bernstein02}:
\begin{align}
(\gamma_{+},\gamma_{\times}) = \frac{1}{2\mathcal{R}}\langle (e_{+},e_{\times}\rangle.
\end{align}
The observed (projected) shape of a galaxy has contributions from both intrinsic shape ($\gamma^{I}$) and gravitational shear ($\gamma^{G}$). Since we are concerned with coherent alignment effects that will contribute to the ensemble shape average, in the remainder of this work, we will use shear and ellipticity interchangeably, denoting both with $\gamma$. For small lensing shears, these contributions add: $\gamma^{\rm obs}=\gamma^{G}+\gamma^{I}$. The intrinsic galaxy shapes will in general include both a random ``shape noise'' and a coherent intrinsic alignment. The lensing contribution to the observed shape is typically only $\sim1\%$. 

In configuration space, the galaxy density-weighted shape field (see section~\ref{sec:dens_weight}) is the product of the galaxy density and intrinsic shape fields: $\tilde{\gamma}_{(+,\times)} = (1+\delta_g)\gamma_{(+,\times)}$, where we use the standard definition of the overdensity field: $\delta = \rho/\bar{\rho}-1$. In Fourier space, the natural decomposition of this shape field is into $E$ (curl-free) and $B$ (divergence-free) modes. $E$ and $B$ modes can be thought of as measuring the $+$ and $\times$ components, with respect to the direction of $\mathbf{k}$ (e.g.\ \cite{kamionkowski98}). 
\begin{align}
\label{eq:E_B_defined}
\tilde{\gamma}_E(\mathbf{k}) &= f_{E}(\mathbf{k})\tilde{\gamma}_{+}({\bf k}) + f_{B}(\mathbf{k})\tilde{\gamma}_{\times}({\bf k}),\notag\\
\tilde{\gamma}_B(\mathbf{k}) &= -f_{B}(\mathbf{k})\tilde{\gamma}_{+}({\bf k}) + f_{E}(\mathbf{k})\tilde{\gamma}_{\times}({\bf k}),
\end{align}
where we have defined the angular operators:
\begin{align}
f_E(\mathbf{k})&=\frac{k_x^2-k_y^2}{\kappa^2},\notag\\
f_B(\mathbf{k})&=\frac{2k_x k_y}{\kappa^2},
\end{align}
for $\kappa^2=k_x^2+k_y^2$. We work in the plane-parallel approximation and define the $\hat{z}$-direction to be along the line-of-sight. Because the decomposition into $E$ and $B$ modes is invariant under rotations on the plane of the sky, we can consider $k_y=0$ modes only, without loss of generality. In terms of these Fourier space fields and the galaxy density field $\delta_g(\mathbf{k})$, we can define the following power spectra: $P_{gE}(\mathbf{k})$, $P_{EE}(\mathbf{k})$, and $P_{BB}(\mathbf{k})$. By symmetry under parity, $P_{gB}(\mathbf{k})$ and $P_{EB}(\mathbf{k})$ must be zero. Similarly, the configuration space correlation functions $\xi_{g\times}=\xi_{+\times}=0$.

From Equation~\ref{eq:E_B_defined}, the relevant power spectra are:
\begin{align}
\label{eq:power_spectra_defined}
(2\pi)^3\delta(\mathbf{k}+\mathbf{k'})P_{gE}(k,\theta_k)& = \langle \delta_g (\mathbf{k}) | \tilde{\gamma}_{E}(\mathbf{k'})\rangle = \langle \delta_g (\mathbf{k}) | \tilde{\gamma}_{+}(\mathbf{k'})\rangle_{k_y=0}, \notag \\
(2\pi)^3\delta(\mathbf{k}+\mathbf{k'})P_{EE}(k,\theta_k)&= \langle \tilde{\gamma}_{E} (\mathbf{k}) | \tilde{\gamma}_{E}(\mathbf{k'})\rangle = \langle\tilde{\gamma}_{+}(\mathbf{k}) | \tilde{\gamma}_{+}(\mathbf{k'})\rangle_{k_y=0}, \notag \\
(2\pi)^3\delta(\mathbf{k}+\mathbf{k'})P_{BB}(k,\theta_k)&= \langle \tilde{\gamma}_{B} (\mathbf{k}) | \tilde{\gamma}_{B}(\mathbf{k'})\rangle = \langle\tilde{\gamma}_{\times}(\mathbf{k}) | \tilde{\gamma}_{\times}(\mathbf{k'})\rangle_{k_y=0},
\end{align}
where the power spectra depend on the amplitude $k$ as well as $\theta_k$, the angle between $\hat{k}$ and the line-of-sight. In general, the two galaxy samples being correlated need not be same, particularly in the case of density-shape cross-correlations (e.g., \cite{singh14})

Correlation functions in configuration space are defined for galaxy shape components measured with respect to the projected separation axis between pairs of objects \cite{kamionkowski98}. In terms of the power spectra defined in eq.~\ref{eq:power_spectra_defined}:
\begin{align}
\langle \delta_g | \tilde{\gamma}_{+}\rangle_{\mathbf{r}} = \xi_{g+}(r_p,\Pi)&=\int \frac{d^3k}{(2\pi)^3} e^{i\mathbf{k}\cdot\mathbf{r}}P_{gE}(k,\theta_k)\cos(2\phi_k),\\
\langle \tilde{\gamma}_{+} | \tilde{\gamma}_{+}\rangle_{\mathbf{r}} = \xi_{++}(r_p,\Pi)&=\int \frac{d^3k}{(2\pi)^3} e^{i\mathbf{k}\cdot\mathbf{r}}\left(P_{EE}(k,\theta_k)\cos^2(2\phi_k) + P_{BB}(k,\theta_k)\sin^2(2\phi_k)\right),\notag\\
\langle \tilde{\gamma}_{\times} | \tilde{\gamma}_{\times}\rangle_{\mathbf{r}} = \xi_{\times\times}(r_p,\Pi)&=\int \frac{d^3k}{(2\pi)^3} e^{i\mathbf{k}\cdot\mathbf{r}}\left(P_{EE}(k,\theta_k)\sin^2(2\phi_k) + P_{BB}(k,\theta_k)\cos^2(2\phi_k)\right),\notag
\end{align}
where $r_p$ is projected (transverse) separation, and $\Pi$ is line-of-sight separation. Switching to cylindrical coordinates and performing the $\phi_k$ part of the integral yields the standard relations in terms of Bessel functions $J_n(x)$:
\begin{align}
\label{eq:xi_cylin}
\xi_{g+}(r_p,\Pi)&=\frac{-2}{(2\pi)^2} \int\limits_{0}^{\infty} dk_z d\kappa \frac{\kappa^3}{k^2} \cos(k_z \Pi) J_2(\kappa r_p) P_ {gE}(k),\\
\xi_{(++,\times\times)}(r_p,\Pi)&=\frac{1}{(2\pi)^2} \int\limits_{0}^{\infty} dk_z d\kappa \kappa \cos(k_z \Pi) \notag\\
&\times\left[\left(J_0(\kappa r_p)\pm J_4(\kappa r_p)\right) P_ {EE}(k,\theta_k)+\left(J_0(\kappa r_p)\mp J_4(\kappa r_p)\right) P_ {BB}(k,\theta_k)\right].\notag
\end{align}
The wavevector has been decomposed into components along the line-of-sight ($k_z=k\cos\theta_k$) and on the sky ($\kappa=k\sin\theta_k$). The top line of eq.~\ref{eq:xi_cylin} uses $P_{gE}(k,\theta_k)=\sin^2(\theta_k)P_{gE}(k)$. The $\theta_k$-dependence for $P_{EE}$ and $P_{BB}$ is more complicated, and this dependence is left in the definitions of the power spectra.

The projected correlation functions are defined as $w(r_p)=\int\limits_{-\Pi_{\rm max}}^{\Pi_{\rm max}} d\Pi\, \xi(r_p,\Pi)$, yielding:
\begin{align}
\label{eq:w_general}
w_{g+}(r_p) &= \frac{-4}{(2\pi)^2} \int\limits_{0}^{\infty} dk_z d\kappa \frac{\kappa^3}{k^2} \frac{\sin(k_z \Pi_{\rm max})}{k_z} J_2(\kappa r_p) P_{gE}(k)\\
w_{(++,\times\times)}(r_p)&=\frac{2}{(2\pi)^2} \int\limits_{0}^{\infty} dk_z d\kappa \kappa \frac{\sin(k_z \Pi_{\rm max})}{k_z} \notag\\
&~~\times\left[\left(J_0(\kappa r_p)\pm J_4(\kappa r_p)\right) P_ {EE}(k,\theta_k)+\left(J_0(\kappa r_p)\mp J_4(\kappa r_p)\right) P_ {BB}(k,\theta_k)\right].\notag
\end{align}
For separations $r_p \ll \Pi_{\rm max}$, only $k_z=0$ modes contribute (the Limber approximation):
\begin{align}
\label{eq:w_limber}
w_{g+}(r_p) &\approx \frac{-2\pi}{(2\pi)^2} \int\limits_{0}^{\infty} d\kappa \kappa J_2(\kappa r_p) P_ {gE}(\kappa),\\
w_{(++,\times\times)}(r_p)&\approx \frac{\pi}{(2\pi)^2} \int\limits_{0}^{\infty} d\kappa \kappa \left[\left(J_0(\kappa r_p)\pm J_4(\kappa r_p)\right) P_ {EE}(\kappa)+\left(J_0(\kappa r_p)\mp J_4(\kappa r_p)\right) P_ {BB}(\kappa)\right].\notag
\end{align}
This result can be seen from taking the limit $\Pi_{\rm max} \rightarrow \infty$ in eq.~\ref{eq:w_general}. 

The $E$ and $B$ mode correlation functions can be written as the Fourier transform of the corresponding power spectrum:
\begin{align}
\xi_{(EE,BB)}=\int \frac{d^3k}{(2\pi)^3} e^{i\mathbf{k}\cdot\mathbf{r}}P_{(EE,BB)}(k) .
\end{align}
Equivalently, these correlations can be written as non-local functions of $\xi_{++}$ and $\xi_{\times\times}$ \cite{crittenden02,blazek11}.

\section{Tidal alignment and nonlinear contributions}
\label{sec:tidal_alignment}
The ``tidal alignment'' model (sometimes referred to as ``linear alignment'') posits that intrinsic galaxy ellipticity has a linear dependence on the tidal field \cite{hirata04}:

\begin{align}
\label{eq:ta_model}
\gamma^I_{(+,\times)}=-\frac{C_1}{4\pi G}(\nabla_x^2-\nabla_y^2,2\nabla_x\nabla_y)S[\Psi_{\rm IA}],
\end{align}
where $\Psi$ is the gravitational potential and $C_1$ parameterizes the strength of the alignment, with sign convention such that positive (negative) $C_1$ corresponds to preferential galaxy alignment along the stretching (compressing) axis of the tidal field.  Positive $C_1$ corresponds to the expected alignment behavior and yields an anti-correlation between the intrinsic ellipticity of a foreground galaxy and the gravitational shear of a background galaxy. $S$ is a filter that smooths fluctuations on halo or galactic scales - see section~\ref{sec:smoothing} for further discussion on the effect of smoothing. The gravitational potential is related to the density field through the Poisson equation in Fourier space (valid on sub-horizon scales, $k\gg cH_0$):
\begin{align}
\label{f}
\Psi_{\rm IA}(\mathbf{k},z_{\rm IA}) = -4\pi G \rho_{m,0}(1+z_{\rm IA})k^{-2}\delta(\mathbf{k},z_{\rm IA}),
\end{align}
where $\rho_{m,0}$ is the matter density today. As discussed above, this type of alignment is most likely to arise in elliptical galaxies for which angular momentum does not play a significant role in determining shape and orientation.

We can write the tidal alignment model in Fourier space:
\begin{align}
\label{eq:ta_fourier}
\gamma^{\rm IA}_{(+,\times)}(\mathbf{k},z) &= -C_1(z) \rho_{m,0}(1+z_{\rm IA})f_{(+,\times)}(\mathbf{k})\delta(\mathbf{k},z_{\rm IA}),
\end{align}
where we define the tidal operator to include the relevant smoothing filter:
\begin{align}
\label{eq:fpluscross}
f_{(+,\times)}(\mathbf{k})\equiv \sin^2(\theta_k)f_{(E,B)}(\mathbf{k})S(k).
\end{align}
The amplitude $C_1$ will generally be a function of galaxy properties, such as color and luminosity, as well as observed redshift, reflecting potential IA evolution. The smoothing function $S(k)$ corresponds to convolution by a smoothing kernel in configuration space.

\subsection{Nonlinear contributions}
The tidal alignment ansatz assumes that the intrinsic ellipticity is a purely linear function of the tidal field, leaving three sources of nonlinearity which must be simultaneously considered: (1) nonlinear evolution of the density field; (2) weighting by shape tracer density; (3) nonlinear galaxy biasing. These effects are described below.

\subsubsection{Nonlinear density evolution}
We use the methods of standard perturbation theory (SPT) to expand the nonlinear density field in terms of the linear density field (see, e.g., \cite{bernardeau02}). We introduce the notation $\delta^{(n)}$ to denote a convolution of $n$ linear density fields with the appropriate gravity kernel capturing structure growth (see appendix~\ref{sec:app_SPTcalc}.) The nonlinear density field can then be written:
\begin{align}
\delta = \delta^{(1)} + \delta^{(2)} + \delta^{(3)} + \cdots
\end{align}
The linear density field is Gaussian, and by Wick's theorem, correlations of an odd number of Gaussian fields are zero. Thus, all correlations are expressed in even powers of the linear density field, or equivalently, in an expansion of the linear power spectrum $\Pl$.

\subsubsection{Shape tracer density weighting}
\label{sec:dens_weight}
Although the tidal alignment model makes a prediction for the (correlated) intrinsic shape field as a function of the tidal field (eq.~\ref{eq:ta_model}), the observable quantity that contaminates the weak lensing signal is the shape field weighted by the density of the galaxies used as tracers\footnote{This situation is completely analogous to observations of the cosmological velocity field.}: $\tilde{\gamma}(\mathbf{x}) = (1+\delta_g(\mathbf{x}))\gamma(\mathbf{x})$. Density weighting arises because we cannot sample the intrinsic ellipticity field in an unbiased way, but rather only where luminous tracers form. In Fourier space, density weighting results in a convolution (denoted $a \circ b$): $\tilde{\gamma}(\mathbf{k}) = \gamma(\mathbf{k})+\delta_g(\mathbf{k}) \circ \gamma(\mathbf{k})$. This weighting has important observational consequences. For instance, although the tidal alignment model produces a pure $E$-mode intrinsic shape field (even when including nonlinear evolution), density weighting produces a $B$-mode contribution. Density weighting also increases measured intrinsic shape correlations on small scales as the contribution of galaxy clustering becomes significant.

Finally, density weighting contributes a bias-dependence to the IA amplitude on large scales. As seen in appendix~\ref{sec:app_SPTcalc}, density weighting produces a term of the form $(58/105)b_1^2 \sigma_S^2 \Pl$, where $\sigma_S^2=\int\frac{d^3q}{(2\pi)^3} \Pl(q) S(q)$. Due to mode coupling, regions with a high density of shape tracers also have large tidal field fluctuations, with an increasing effect for larger tracer bias. Density weighting thus yields a bias-dependent enhancement in the observed shape correlations, even if the underlying response to the tidal field is comparable for different tracers. This enhancement is present even on large scales ($k \rightarrow k_{\rm lin}$), where linear theory should be valid, leading to a bias-dependent effective value of $C_1$:
\begin{align}
P_{gE}(k_{\rm lin}) \propto C_1\left(1+\frac{58}{105}b_1 \sigma^2_S\right)\Pl , \notag\\
P_{EE} (k_{\rm lin}) \propto C_1^2\left(1+2\frac{58}{105}b_1 \sigma^2_S\right)\Pl .
\end{align}
It is this effective value of $C_1$ that is actually measured with large-scale intrinsic shape correlations. This effect could be largely responsible for the luminosity dependence observed in shape correlations (e.g.\ \cite{hirata07,faltenbacher09,joachimi11,singh14}), and the trend with $b_1$ it predicts is consistent with the measurements of \cite{singh14}. However, this effect may be partly degenerate with other dependence of $C_1$ on galaxy properties, including galaxy-halo misalignment \cite{okumura09b,faltenbacher09,blazek11}. The value of $\sigma_S^2$ depends explicitly on the small-scale cut-off in tidal fluctuations. Moreover, since the correction is generally large, higher-loop contributions may be important. Thus, although it is possibly to calculate $\sigma^2_S$ under a particular smoothing procedure, it may be preferable to instead treat $\sigma^2_S$ as an additional free parameter of the model.

\subsubsection{Nonlinear galaxy bias}
We include nonlinear bias contributions to the galaxy density field. In Fourier space:
\begin{align}
\label{eq:nonlin_bias}
\delta_g &= b_1 \delta + b_2 (\delta \circ \delta) -b_2 \sigma^2+ b_3 (\delta \circ \delta \circ \delta) + \cdots.
 \end{align}
where $\delta$ is the nonlinear density field, and $\sigma^2 \equiv \langle \delta^2 \rangle$. A more general bias expansion at this order would include the nonlocal tidal terms $b_s$ and $b_{3,{\rm nl}}$ (e.g.\ \cite{baldauf12,saito14}). We leave consideration of these tidal bias contributions for future work.

\subsection{Alignment epoch}
The redshift at which the alignment is set, $z_{\rm IA}$, and the subsequent IA evolution, are determined by the astrophysical processes involved in galaxy formation and evolution. It is sometimes assumed that $z_{\rm IA}$ is during matter domination when a halo first forms (``primordial alignment''). However, late-time accretion and mergers could have a significant impact on IA, in which case the relevant $z_{\rm IA}$ could be closer to the observed redshift, $z_{\rm obs}$. The limiting case of $z_{\rm IA}=z_{\rm obs}$ is the ``instantaneous alignment'' scenario. For further discussion on this topic, see \cite{blazek11,kirk12,camelio15}.

At leading order, the impact of alignment epoch can be absorbed into a redshift-dependent amplitude, $A(z)$, and eq.~\ref{eq:ta_fourier} can be written:
\begin{align}
\label{eq:ta_fourier2}
\gamma^{\rm IA}_{(+,\times)}(\mathbf{k},z) &= A(z) f_{(+,\times)}(\mathbf{k})\delta(\mathbf{k},z).
\end{align}
In the case of instantaneous alignment ($z_{\rm IA}=z_{\rm obs}$):
\begin{align}
A(z)=-C_1(z)\rho_{m,0}(1+z).
\end{align}
For earlier epochs of alignment, the linear order evolution of the density field evolution can be absorbed into the growth factor $D(z)$:
\begin{align}
\label{eq:general_A}
A(z)=-C_1(z)\rho_{m,0}(1+z_{\rm IA})\frac{D(z_{\rm IA})}{D(z)}.
\end{align}
In the case of primordial alignment during matter domination, the IA field is insensitive (at linear order) to $z_{\rm IA}$:
\begin{align}
\label{eq:prim_A}
A(z)=\frac{-C_1(z)\rho_{m,0}(1+z)}{\bar{D}(z)}\,
\end{align}
where $\bar{D}=(1+z)D(z)$ is the rescaled growth factor, normalized such that $\bar{D}(z)=1$ during matter domination. In these expressions, we have separated redshift dependence due to the growth of structure from possibly evolution in the response to the tidal field, captured in $C_1(z)$. 

Once nonlinear evolution of the density field is included, eq.~\ref{eq:ta_fourier2} is only exact in the case of instantaneous alignment. Equations \ref{eq:general_A} and \ref{eq:prim_A} provide only approximate amplitude scalings when $z_{\rm IA} > z_{\rm obs}$, since the choice of $z_{\rm IA}$ will affect both the overall amplitude of the IA field as well as its scale dependence (via nonlinear evolution). In this case, nonlinear predictions for the observed IA correlations must be treated carefully. For instance, if the intrinsic shapes and alignments of galaxies are set at the redshift of initial formation, $z_P$, the tidal field should be evaluated at that redshift. A model is then required to describe the subsequent evolution of halo and galaxy alignments. However, the relevant shape and density tracers fields should still be evaluated at $z_{\rm obs}$.

Moreover, even if the alignment field is set at $z_P$ and simply ``passively'' evolves to $z_{\rm obs}$, it will be effectively transformed by the motion of the halos/galaxies on which it was initially imprinted. In our Eulerian approach, the advection of modes due to fluctuations on larger scales will suppress correlations of density (or tidal) fields at different redshifts, which is non-physical in the case where primordial alignments are ``carried'' with the moving galaxies. At one loop order, this effect appears as a relative suppression of the mode-mixing term compared to the propagator contribution.

One simple approach to modeling the evolution of IA set at high $z_{\rm IA}$ has been to take the geometric mean between the power spectra evaluated at the two relevant redshifts, $z_{\rm IA}$ and $z_{\rm obs}$ (e.g. \cite{kirk12}). While such an approach avoids the non-physical suppression that arises in the Eulerian treatment, it also fails to fully treat advection effects and should thus be treated with caution. In this work, we adopt the instantaneous alignment ansatz, for which Eulerian SPT is well-suited. We leave a more thorough treatment of nonlinear advection of the IA field for future work.

\subsection{Smoothing the tidal field}
\label{sec:smoothing}
Tidal field fluctuations on sufficiently small scales should not have a significant impact on the observed IA of halos/galaxies, and it is thus common to smooth the tidal field (e.g. \cite{hirata04,chisari13}). The correct prescription for smoothing is not settled - from a physical standpoint, smoothing reflects the finite size of the halos/galaxies and the regions from which they formed. Alternatively, the smoothing can be viewed as an element in any effective theory, in which small-scale physics is absorbed into the measured parameter values.

We treat the smoothing filter as a multiplicative function in Fourier space. Smoothing is applied to the tidal field, via a redefinition of the IA operator (eq.~\ref{eq:fpluscross}), before the convolution integrals are performed. We choose the form:
\begin{align}
S(k)&=e^{-\left(\frac{k}{k_{\rm sm}}\right)^2},\notag\\
k_{\rm sm}&=\alpha \frac{2\pi}{R_L},
\end{align}
where $R_L$ is the Lagrangian radius of the halo, and $\alpha$ is an order-unity factor to capture the uncertainty in the IA response to smoothing. The Lagrangian radius of a halo with mass $M_{\rm halo}$ is defined as the radius of a sphere in a homogenous universe that contains the halo's mass:
\begin{align}
M_{\rm halo}=\frac{4\pi}{3} \bar{\rho}_m R_L^3 .
\end{align}
In the tidal alignment picture, it is often assumed that halo shapes are determined by the external tidal field during collapse, making the Lagrangian radius a natural scale for smoothing. Fluctuations on smaller scales may significantly influence the actual shapes of the inner regions of halos (and the galaxies residing there), but should not contribute to correlations on large scales \cite{blazek11}. For the typical luminous red galaxies (LRGs) used for comparison in this work, the halo mass is $M_{\rm halo} \sim 3 \times 10^{13}~h^{-1}M_{\odot}$. We can write $R_L$, scaled by typical parameter values:
\begin{align}
R_L= (4.41~h^{-1}{\rm Mpc})\times\left( \frac{M_{\rm halo}}{ 3 \times 10^{13}~h^{-1} M_{\odot}} \right)^{1/3}\left( \frac{\Omega_{\rm m,0}}{0.3}\right)^{-1/3} .
\end{align}
Note that in comoving coordinates, $R_L$ is constant with redshift, since the product of volume and density remains fixed. In this work, we choose $k_{\rm sm}=1.0$ \hMpcinv. The mass dependence of the Lagrangian radius, and this smoothing scale, is fairly weak.

The tidal field on smaller scales may have an important role in determining actual galaxy shapes and orientations, particularly when considering the central baryonic component, but these effects simply modulate the large-scale correlations, analogous to how nonlinear galaxy formation physics modulates the large-scale galaxy bias. The physics of IA may actually yield a different smoothing scale than the one assumed here. Indeed, comparing predictions with different smoothing scales to measured IA provides one way to probe the underlying effects of tidal field fluctuations on different scales.

\subsection{Redshift-space distortions}
The peculiar velocities of galaxies lead to shifts in their apparent distance along the line-of-sight. On linear scales, the effect of these ``redshift-space distortions'' (RSDs) on the galaxy density field are well-described by the Kaiser formula \cite{kaiser87}:
\begin{align}
\delta^s_g(k)=(b_1+f\mu^2)\delta^r(k),
\end{align}
where $f=d \ln D/d \ln a$ is the logarithmic growth rate, $\mu=\hat{k}\cdot\hat{z}$, and ``r'' and ``s'' denote quantities in real and redshift space, respectively. Although the apparent line-of-sight positions of galaxies are affected by RSD, their observed shapes are not. Thus, at leading order: $\gamma^s=\gamma^r$. The density-weighted intrinsic shape field in redshift space ($\tilde{\gamma}^s$) is affected by the shift in observed positions of shape tracers as well as higher-order effects on $\gamma^s$. For further discussion, see the appendix of \cite{singh14}.

For the projected statistics considered here, only RSDs on linear scales will contribute significantly, suggesting that we can consider only the linear contribution to $\tilde{\gamma}^s$, for which the effect of RSDs can be ignored. In this case:
\begin{align}
\label{eq:IA_RSD1}
P^s_{gE}&=\left(1+\frac{f}{b_1}\mu^2\right)P^r_{gE},\notag\\
P^s_{EE,BB}&=P^r_{EE,BB}.
\end{align} 
We use this correction in the results presented below. Note that the treatment of RSDs in \cite{blazek11,chisari13} assigns a linear-order effect to both $\delta_g$ and $\tilde{\gamma}$, yielding a small misestimate of the IA correlation function. For instance, for a projection length of $\Pi_{\rm max} = 80$\hMpc, the RSD treatment of eq.~\ref{eq:IA_RSD1} yields to a $\sim2\%$ correction to the real-space $w_{g+}$ at $r_p=50 h^{-1}{\rm Mpc}$, compared to a $\sim4\%$ correction in \cite{blazek11,chisari13}. At smaller projected separation, RSDs rapidly become negligible.

The linear theory result of eq,~\ref{eq:IA_RSD1} neglects an $\mathcal{O}(\Plsq)$ contribution from density weighting that yields a term proportional to $\Pl$ on large scales (see section~\ref{sec:dens_weight}). This term is present in both $P^s_{gE}$ and $P^s_{EE}$ and will contain an additional Kaiser-like enhancement factor. However, given the small impact of RSDs on the scales of interest, we do not consider this contribution here.
\subsection{Combining all contributions}
\label{sec:combining}
In this section, we present the key results from combining all relevant contributions to tidal alignment. Details can be found in appendix~\ref{sec:app_SPTcalc}. Using SPT at one loop, we calculate the relevant power spectra, defined in eq.~\ref{eq:power_spectra_defined}. We then transform to the correlation functions (eq.~\ref{eq:w_general}) to compare with measurements.

Combining eqs.~\ref{eq:nonlin_bias}-\ref{eq:ta_fourier2} and keeping terms up to $\mathcal{O}(\delta^4)$, the density weighted intrinsic shape field is:
\begin{align}
\tilde{\gamma}_{i}(k,z) &=A(z) \left(  (1-b_2\sigma^2) f_i\delta + b_1 \delta \circ (f_i\delta) + b_2 (\delta \circ \delta) \circ (f_i\delta) + b_3 (\delta \circ \delta \circ \delta) \circ (f_i\delta) \right)
\end{align}
Collecting all contributions to the cross-correlation between galaxy density and shape that are not zero by symmetry, we find:
\begin{align}
\label{eq:PgE_corrs}
\langle \delta_g | \tilde{\gamma}_{+}\rangle = A(z) &\times\Big[
  b_1 f_+\langle \delta | \delta \rangle + b_1^2 \langle \delta | \delta \circ (f_+\delta)\rangle +b_2 f_+ \langle \delta \circ \delta | \delta \rangle\notag \\
 &+b_1b_2 \langle \delta \circ \delta | \delta \circ (f_+ \delta) \rangle_c
 + \mathcal{O}(\delta_{L}^6)\Big],
\end{align}
where the $c$ subscript indicates connected correlators, with two or more branches between the two points. Similarly, the intrinsic shape auto-correlations can be calculated, up to one loop:
\begin{align}
\label{eq:Pauto_corrs}
\langle \tilde{\gamma}_{i} | \tilde{\gamma}_{i} \rangle = A^2(z)&\times\Big[
f^2_i\langle \delta | \delta \rangle + 2 b_1 \langle \delta | \delta \circ (f_i\delta)\rangle +b_1^2\langle \delta \circ (f_i\delta)| \delta \circ (f_i\delta)\rangle_c + \mathcal{O}(\delta_{L}^6)\Big],
\end{align}
We define power spectra corresponding to the correlators in equations~\ref{eq:PgE_corrs}-\ref{eq:Pauto_corrs}, recalling that we can evaluate these correlations at $k_y=0$:
\begin{align}
\label{eq:PgE1}
P_{gE}(\mathbf{k})&= A(z) \sin^2(\theta_k)\left[b_1 P_{0|S}+b_1^2P_{0|0E} + b_2 P_{00|S} +b_1 b_2 P_{00|0E} + \mathcal{O}(P_L^3)\right],\\
\label{eq:PEE1}
P_{EE}(\mathbf{k}) &=A^2(z)\times \Big[\sin^4(\theta_k)P_{S|S}
+2b_1 \sin^4(\theta_k) P_{S|0E}
+b_1^2 P_{0E|0E}(\theta_k) + \mathcal{O}(P_L^3)],\\
\label{eq:PBB1}
P_{BB}(\mathbf{k}) &=A^2(z)\times \Big[b_1^2 P_{0B|0B}(\theta_k)
+ \mathcal{O}(P_L^3)\Big].
\end{align}
The subscripts in these expressions indicate the fields being correlated: density field (0), smoothed density field ($S$), or components of the smoothed tidal field ($E$ or $B$). For instance, $P_{0|S}$ is the nonlinear matter power spectrum in which one power of the density field is smoothed at the relevant IA scale. Dependence on $\theta_k$ is explicitly indicated. For the exact form of these terms in SPT, see appendix~\ref{sec:app_SPTcalc}. When the projection length is larger than the scale of interest, the Limber approximation can be used, and only $k_z=0$ modes will contribute, corresponding to $\sin^2(\theta_k)=1$.

Although we arrived at equations~\ref{eq:PgE1}-\ref{eq:PBB1} through an SPT expansion, each term can be evaluated to arbitrary precision. For instance, SPT is known to provide a comparatively poor fit to the nonlinear matter power spectrum, $\langle \delta | \delta \rangle$ (e.g.\ \cite{carlson09}). More accurate techniques are available to calculate this term (e.g.\ a fitting formula such as Halofit \cite{smith03} or a simulation emulator \cite{heitmann14}). In the following results, we use a Halofit nonlinear power spectrum for $P_{0|0}$, $P_{0|S}$, and $P_{S|S}$. 

\begin{figure}[h!]
\begin{center}
\resizebox{5.5in}{!}{
\includegraphics{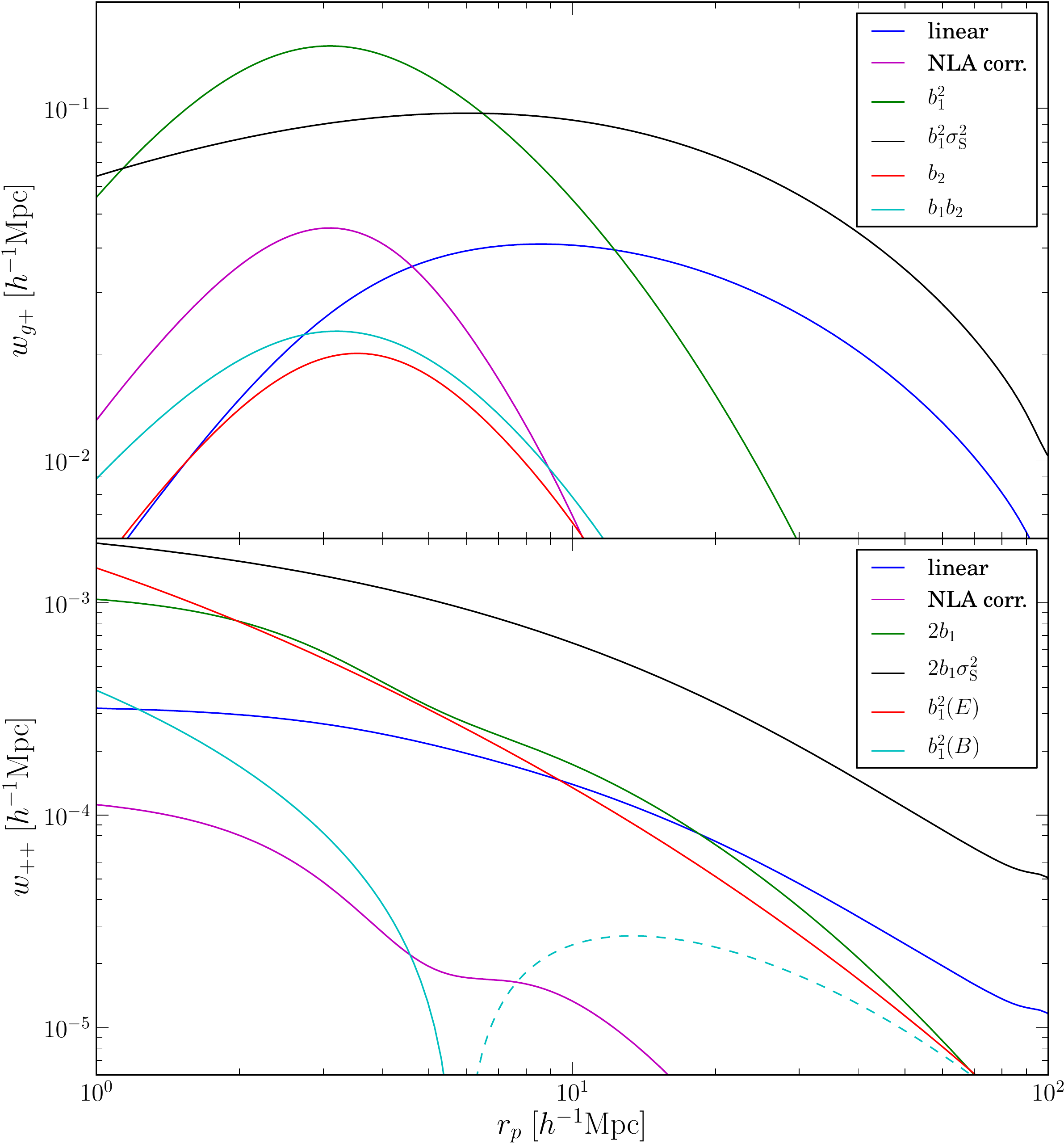}
}
\end{center}
\caption{The linear theory result, NLA correction, and all additional $\mathcal{O}(\Pl^2)$ corrections are shown for $w_{g+}$ ({\it top panel}) and $w_{++}$ ({\it bottom panel}), with $b_1=2.1$ and $b_2=0.5$. Overall normalization for both statistics, determined by $C_1$, is chosen to match the LOWZ $w_{g+}$ measurement of \cite{singh14}, while the value of $\sigma_S^2$ is calculated using the smoothing discussed in section~\ref{sec:smoothing}, with $k_{\rm sm}=1.0$ \hMpcinv. Contributions are labeled using the associated pre-factors in eqs.~\ref{eq:PgE1}-\ref{eq:PBB1}, and the $\sigma_S^2\Pl$ contribution (black) is separated from the other tracer density weighting terms (green). Note that the ``NLA corr.''\ term shows the difference between the NLA (Halofit) and linear theory predictions. Dashed lines indicate a negative value.}
\label{fig:wcomps}
\end{figure}

Figure~\ref{fig:wcomps} shows the linear and nonlinear contributions to $w_{g+}$ and $w_{++}$ in the tidal alignment model for an LRG-like shape and density tracer sample (e.g.\ \cite{okumura09a,okumura09b}), for which we assume bias values $b_1=2.1$ and $b_2=0.5$. The $\sigma_S^2 \Pl$ contribution from density weighting rescales the linear theory prediction (i.e.\ the effective value of $C_1$ at large scales) - it is shown with an amplitude set by the one-loop prediction (see section~\ref{sec:dens_weight}). While the contributions from nonlinear galaxy bias are subdominant, the effect of density weighting is significant and is larger than the correction from nonlinear dark matter clustering for these highly biased tracers. We thus confirm that the NLA model omits important nonlinear effects.

\subsection{Comparison to observed IA}
\label{sec:data}

We compare predictions of this nonlinear tidal alignment model with two different IA measurements using highly biased tracers with spectroscopic redshifts from the Sloan Digital Sky Survey (SDSS, \cite{york00}). The measurements of \cite{okumura09a,okumura09b} use $\sim 84,000$ luminous red galaxies (LRGs) from SDSS DR6, with spectroscopic redshifts in the range $0.16<z<0.47$ and mean redshift $\bar{z}=0.32$. We also compare with the recent measurements of \cite{singh14} from the BOSS ``LOWZ'' sample from SDSS DR11. Within the redshift range $0.16<z<0.36$ (for which the sample is roughly volume limited), \cite{singh14} use $\sim 160,000$ galaxies with good shape measurements, yielding a mean redshift $\bar{z}=0.28$.

While the galaxy properties in these two samples are similar, the LRG sample used by \cite{okumura09a,okumura09b} includes objects at higher redshift and has a higher average bias than the LOWZ sample of \cite{singh14}. In addition, the LRG measurements of \cite{okumura09a,okumura09b} use isophotal galaxy shapes which have not been corrected for the point-spread function (PSF). Isophotal shapes apply more weight to the outer regions of galaxies, leading to larger measured values of ellipticity and all related statistics.\footnote{R. Mandelbaum, private communication. It remains an open question whether the PSF introduces spurious correlations that contaminate $w_{g+}$ and $w_{++}$.} For the IA autocorrelation, $w_{++}$, the LRG measurements of \cite{okumura09a} were reprocessed by \cite{blazek11} to account for projection effects and the treatment of randoms in the original estimator. The LOWZ $w_{++}$ measurement of \cite{singh14} has significantly lower signal-to-noise. This discrepancy may be due to the use of different shape measurement techniques or the projection procedure used in \cite{blazek11}. See \cite{singh14} for further discussion. 

\begin{figure}[h!]
\begin{center}
\resizebox{5.5in}{!}{
\includegraphics{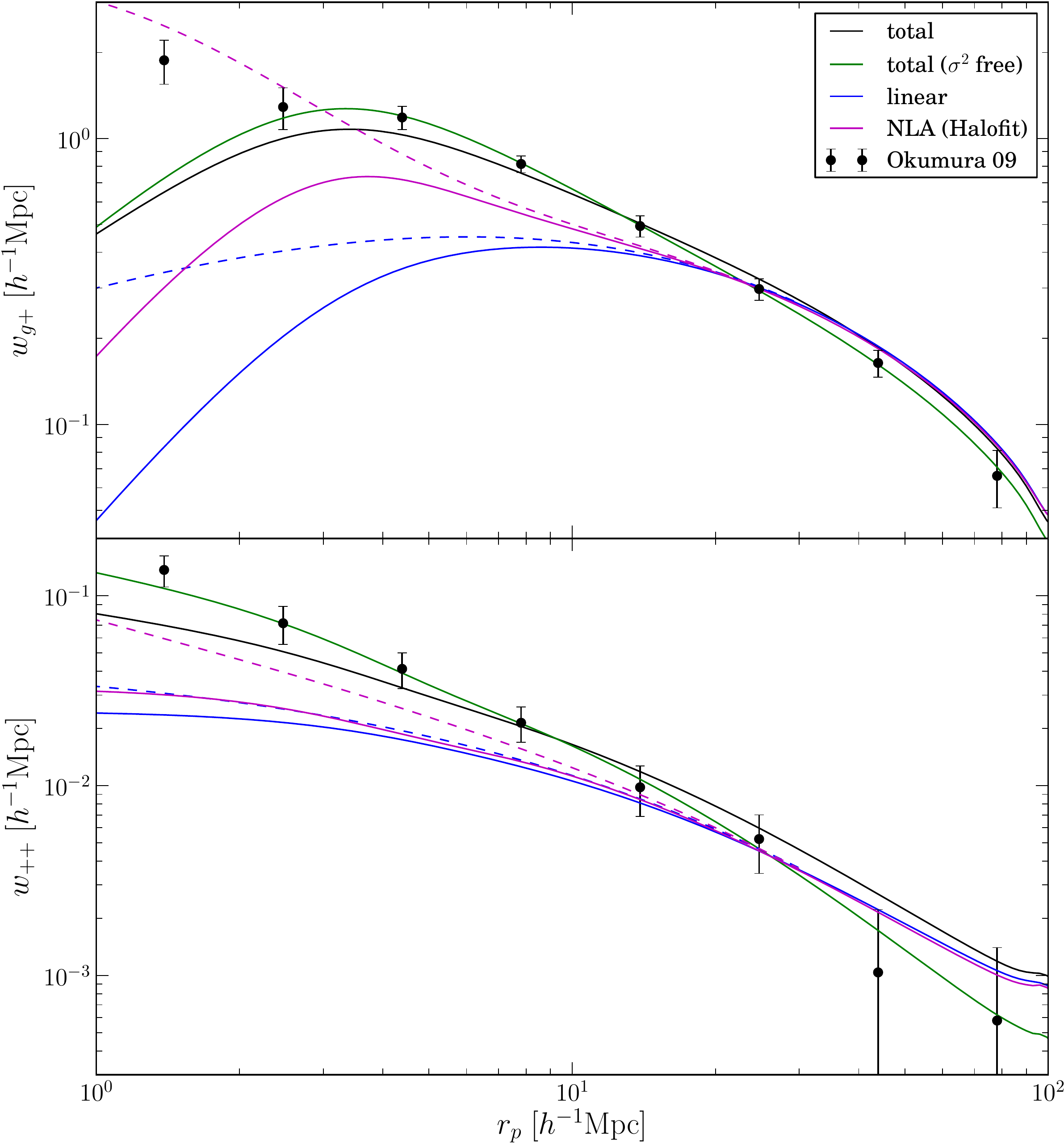}
}
\end{center}
\caption{Data points show the LRG measurements of $w_{g+}$ ({\it top panel}) and $w_{++}$ ({\it bottom panel}) from \cite{okumura09a,okumura09b}. The solid lines show the best fit linear, NLA, and total model predictions, with $b_1=2.1$ and $b_2=0.5$.  For comparison, dashed lines indicate the linear and NLA models with no smoothing of the tidal field. The alignment amplitude and $\sigma^2_S$ is fit separately to $w_{g+}$ and $w_{++}$.}
\label{fig:okumura1}
\end{figure}

\begin{figure}[h!]
\begin{center}
\resizebox{5.5in}{!}{
\includegraphics{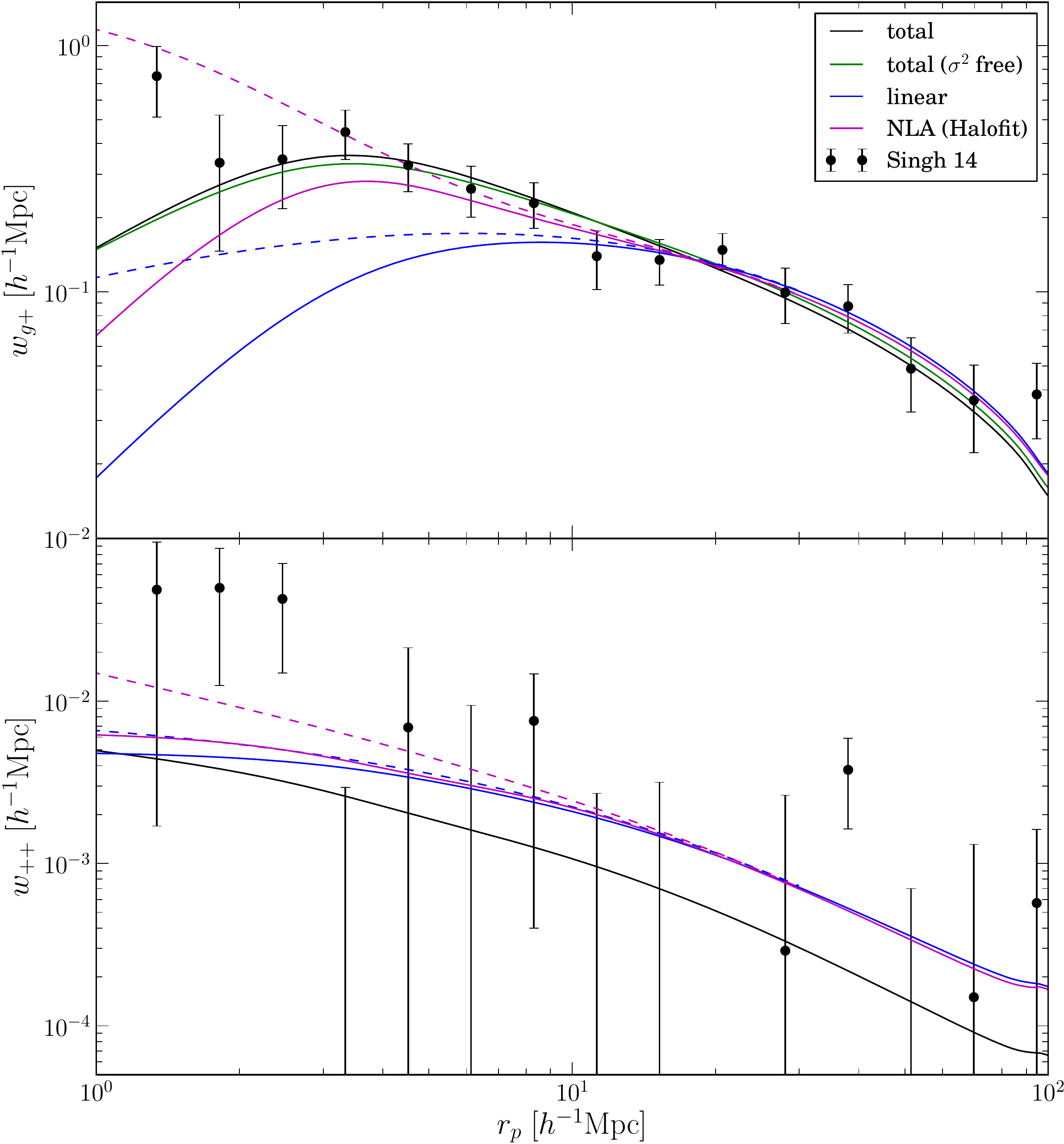}
}
\end{center}
\caption{Data points show the LOWZ measurements of $w_{g+}$ ({\it top panel}) and $w_{++}$ ({\it bottom panel}) from \cite{singh14}. For $w_{g+}$, the solid lines show the best fit linear, NLA, and total model predictions, with $b_1=1.8$ and $b_2=0.5$.  For comparison, dashed lines indicate the linear and NLA models with no smoothing of the tidal field. The linear and NLA models for $w_{++}$ are shown, normalized using the fit to $w_{g+}$ due to low signal-to-noise in $w_{++}$.}
\label{fig:singh1}
\end{figure}

Figures~\ref{fig:okumura1}-\ref{fig:singh1} compare the tidal alignment model with all $\mathcal{O}(\Pl^2)$ corrections with the shape correlation measurements of these two different samples. The alignment amplitude is fit to measurements of $w_{g+}$ above 2 \hMpc~are shown for both fixed $\sigma^2_S$ (calculated using by smoothing the density field as discussed in section~\ref{sec:smoothing}) as well as allowing $\sigma^2_S$ to be a free parameter (constrained to be positive). Also shown are the linear and NLA model predictions, with and without smoothing of the tidal field, fit to measurements above 10 \hMpc. The model is separately fit to the LRG $w_{++}$. As discussed in \cite{blazek11}, the amplitude on large scales is consistent with that of $w_{g+}$. The low signal-to-noise in the LOWZ $w_{++}$ measurement prevents a meaningful model fit. For reference, the linear and NLA models, normalized using the fit to the LOWZ $w_{g+}$, are shown in fig.~\ref{fig:singh1}. In summary, including all nonlinear corrections improves the fit when compared with the linear or NLA models. Allowing $\sigma^2_S$ to be a free parameter further improves the fit, particularly for the LRG $w_{++}$ measurements.

\section{Tidal alignment on small scales}
\label{sec:separation}

The perturbative treatment described in the previous section provides a consistent approach to understanding the contributions to shape correlations in the tidal alignment picture. However, it will necessarily break down on sufficiently small scales. In particular, the tracer clustering (density auto-correlation) on small scales is not well-described in a perturbative framework. In this section, we employ an alternative approach, explicitly separating the effects of galaxy clustering and galaxy alignment and modeling each separately.

Without loss of generality (assuming approximately constant source number density across the projection length, $2\Pi_{\rm max}$), we can divide the shape correlation statistics into contributions from the intrinsic shape field and galaxy clustering:
\begin{align}
\label{eq:sep1}
w_{g+}(r_p) &= \avgplust (w_{gg}(r_p) + 2\Pi_{\rm max}),\notag\\
w_{(++,\times\times)}(r_p) &= \langle {\gamma}_{(+,\times)} {\gamma}_{(+,\times)} \rangle^{\rm trac}_{r_p}(w_{gg}(r_p) + 2\Pi_{\rm max}).
\end{align}
Equation~\ref{eq:sep1} defines the quantities $\avgplust$ and $\langle \gamma_{(+,\times)} \gamma_{(+,\times)} \rangle^{\rm trac}_{r_p}$, which are interpreted as the average alignment between shape and density tracers and between two shape tracers, respectively, at a given projected separation. As this division makes clear, on large scales, the clustering of shape tracers can be ignored ($w_{gg} \ll 2\Pi_{\rm max}$), and shape correlations are determined entirely by these average quantities. On smaller scales, the contribution from tracer clustering can be significant.

\subsection{Modeling $\avgplust$ around halos using tidal alignment}
In this section, we construct a model for $\avgplust$ around biased tracers of dark matter (e.g.\ the correlations relevant to the $w_{g+}$ statistic). Similar arguments apply to modeling $w_{(++,\times\times)}$. The average of the intrinsic shape field at projected separation $r_p$ from a density tracer can be written:
\begin{align}
\avgplus_{r_p} = \frac{1}{2\Pi_{\rm max}}\int\limits_{-\Pi_{\rm max}}^{+\Pi_{\rm max}}d\Pi \,\langle \delta_g | \gamma_+ \rangle_{\mathbf{r}}
\end{align}
Note that this expression is very similar to the definition of $w_{g+}$, except the effect of weighting by the shape tracer density is not present. 

In the tidal alignment framework, the intrinsic shape field in a given location is determined by the local tidal field. The quantity $\avgplus$ is thus determined by the correlation between density tracer positions and the tidal field. Following the results of section~\ref{sec:shape_formalism}, it is clear that $\avgplus$ will essentially be the projected cross-correlation between galaxies (the density tracer) and dark matter, except that the derivatives relating the density to the tidal field will result in a $J_2$ rather than a $J_0$ Hankel transform:
\begin{align}
\avgplus &= \frac{1}{2\Pi_{\rm max}}A(z)w_{g\delta,2}\notag,\\
&\approx \frac{1}{2\Pi_{\rm max}}\frac{A(z)}{(2\pi)} \int\limits_{0}^{\infty} d\kappa \kappa J_2(\kappa r_p) P_ {g\delta}(\kappa),
\end{align}
where the second line assumes the Limber approximation (e.g.\ eq.~\ref{eq:w_limber}), and we have introduced the correlation function notation:
\begin{align}
\xi_{ab,n}(r_p,\Pi)&\equiv\frac{1}{(2\pi)^2} \int\limits_{0}^{\infty} dk_z d\kappa \kappa \cos(k_z \Pi)J_n(\kappa r_p)P_{ab}(k,\theta_k)
\end{align}
for the correlation between quantities $a$ and $b$, weighted by the relevant Bessel function $J_n$ to account for the azimuthal angular dependence (see eq.~\ref{eq:xi_cylin}).

To relate $\avgplus$ to $\avgplust$ we must include the effect of sampling the tidal field at the locations where shape tracers form. As discussed in section~\ref{sec:dens_weight}, in the tidal alignment framework, this effect produces an additional contribution to alignment $\sim b_1 \sigma_S^2\Pl$, which changes the effective large-scale alignment amplitude: $C^{\rm eff}_1 = (1+58/105 b_1 \sigma_S^2)C_1$. We will assume that $\avgplust$ includes this additional term, such that on large scales $\avgplust = (C^{\rm eff}_1/C_1)\avgplus_{r_p}$. The contribution from shape tracer clustering has been explicitly separated into the $w_{gg}$ term in eq.~\ref{eq:sep1}.
\begin{align}
\label{eq:plust}
\avgplust = \avgplus_{r_p} + \frac{A(z)}{2\Pi_{\rm max}}\frac{58}{105} b_1^2\sigma_S^2 w_{\delta\delta,2}^{\rm lin}
\end{align}

We are free to choose a model for $P_{g\delta}$, the cross-power spectrum of galaxies and dark matter. This correlation could be treated perturbatively, as done in the previous section, by including nonlinear galaxy bias. Instead, to describe tidal alignment down to highly non-perturbative scales, we use a Navarro-Frenk-White (NFW) halo density profile \cite{navarro96}:
\begin{align}
\rho(r) = \frac{\rho_s}{(r/r_s)(1+r/r_s)^2},
\end{align}
defined by its scale density ($\rho_s$) and radius ($r_s$). Equivalently, the NFW profile can be described with the halo mass (typically defined at some density threshold) and concentration (the ratio of the profile scale radius to halo radius, defined by the same density threshold). Using galaxy-galaxy lensing, and assuming the mass-concentration relationship of \cite{mandelbaum08b}, \cite{singh14} measure the average LOWZ halo mass to be $M_{180b}=1.51\times10^{13} h^{-1} M_{\odot}$, where $M_{180b}$ denotes the mass enclosed within a region of 180 times the average matter density. To model $P_{g\delta}$, we truncate the NFW profile at $2R_{180b}$, take its Fourier transform, and combine with the linear $P_{g\delta}(k) = b_1\Pl(k)$ to include correlations on larger scales. We note that more sophisticated models for $P_{g\delta}$ could be used within this framework, particularly to describe the transition region between the one- and two-halo regimes. Nevertheless, our approach provides relatively accurate predictions. Rather than apply a smoothing filter to the tidal field, we explicitly separate the alignment behavior within the one-halo regime (discussed below in section~\ref{sec:one_halo}).

Modeling $w_{g+}$ requires a description of both $\avgplust$ and $w_{gg}$. An analytic description of $w_{gg}$ on small scales, particularly the one-halo regime, requires halo modeling specific to the relevant shape and density tracer samples. However, we are also free to use an empirical model (from simulations or observations). In general, the projected clustering signal will be measured to significantly higher precision than the associated shape correlations, and thus in some contexts, using the observed $w_{gg}$ as a component of the intrinsic alignment model is feasible. On small scales, $w_{gg} \gg 2\Pi_{\rm max}$, and thus $w_{g+}\approx \avgplust w_{gg}$. Within the one-halo regime, as we argue below, $\avgplust$ may be approximately constant, leading to $w_{g+} \propto w_{gg}$.

\begin{figure}[h!]
\begin{center}
\resizebox{5.5in}{!}{
\includegraphics{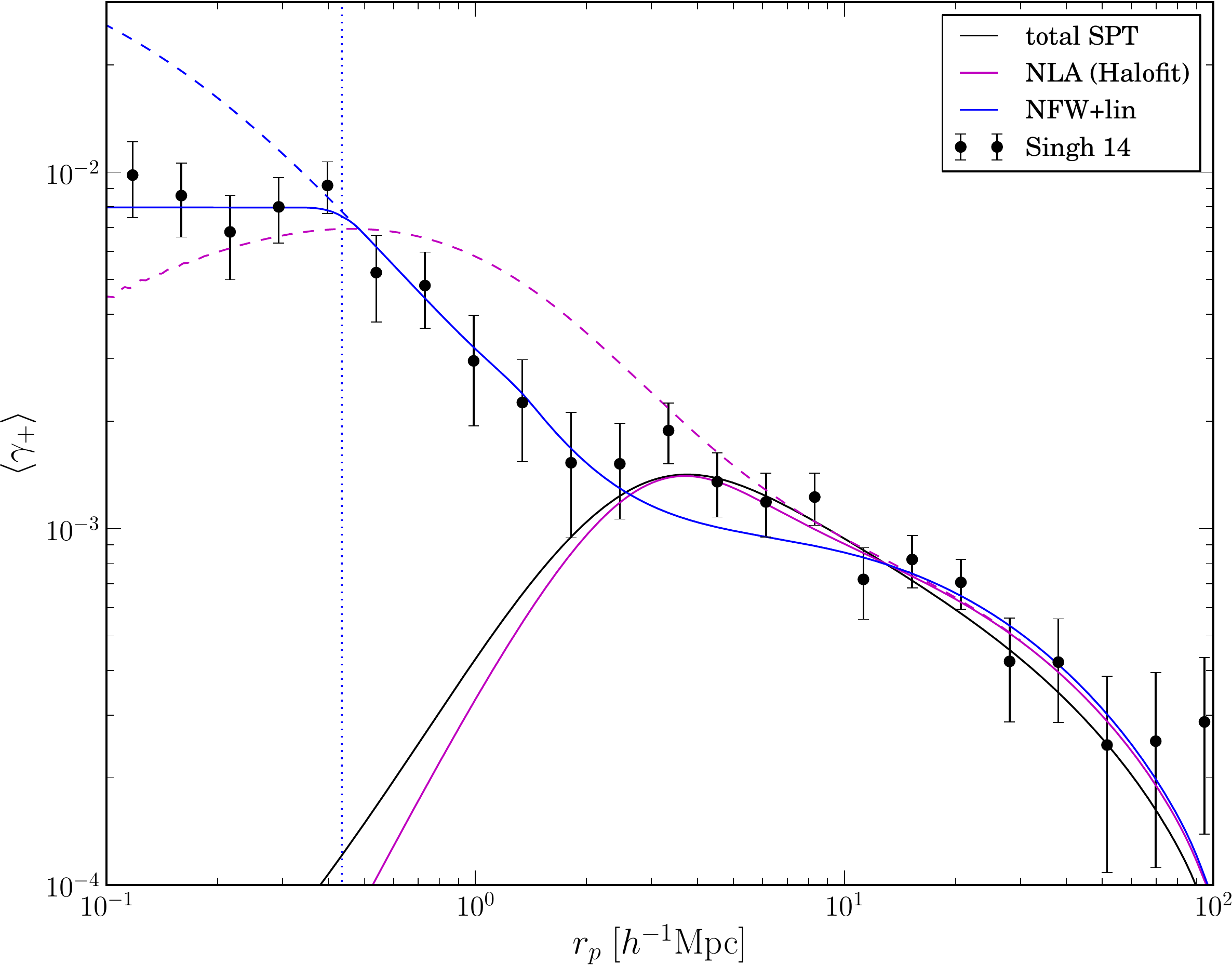}
}
\end{center}
\caption{Model results for $\avgplust$, shown for an NFW profile for $M_{180b}=1.51\times10^{13} h^{-1} M_{\odot}$, is compared with the observational results from the LOWZ sample \cite{singh14}. The dashed blue line shows the continuation of the tidal alignment prediction below $R_{200c}$ (indicated by the vertical line). Predictions from the full SPT model of section~\ref{sec:tidal_alignment}, as well as the NLA model (dashed line indicates no smoothing), are shown for comparison. As discussed in the text, the NLA model does not account for density weighting, and the associated prediction for $\avgplust$ does not include $w_{gg}$.}
\label{fig:avg_IA1}
\end{figure}

Figure~\ref{fig:avg_IA1} compares the result of our model for $\avgplust$ with the LOWZ observational results from \cite{singh14}. With the treatment of the one-halo regime discussed in section~\ref{sec:one_halo}, our model for $\avgplust$ agrees well with the observational results. For comparison, we also show the SPT results of section~\ref{sec:tidal_alignment}. In the case where density weighting is included, we calculate $\avgplust$ by dividing the model prediction for $w_{g+}$ by $(w_{gg} + 2\Pi_{\rm max})$, where we have used Halofit nonlinear matter clustering and linear bias to model $w_{gg}$. For the NLA model, which does not account for source density weighting, we show $w_{g+}$ divided by $2\Pi_{\rm max}$ only.

\subsection{The one-halo regime}
\label{sec:one_halo}

On large scales, in the two-halo regime, alignments largely reflect initial correlations in the tidal field. On these scales, tidal alignment (and perhaps tidal torquing) mechanisms should accurately describe intrinsic shape alignments. In the one-halo regime, however, information of these initial conditions is largely scrambled by nonlinear gravitational processes. We thus expect that these theories will no longer provide a qualitatively correct description of shape alignments, even if a fully non-linear treatment of the tidal field (e.g.\ an NFW halo profile) is used. On these scales, $\avgplust$ should differ from the tidal alignment prediction $\propto w_{g\delta,2}$ made above.

Understanding the alignment of central and satellite objects within the same halo remains a crucial question in both observational and theoretical studies. Within a halo, there are three potential types of shape-position alignment: (1) alignment of central shapes with satellite positions; (2) alignment of satellite shapes with the central position; (3) alignment of satellite shapes with other satellite positions. The alignment of central shapes with satellite positions has been measured (e.g.\ \cite{hao11,singh14}). Since satellites should essentially trace the halo density, this effect is expected in the tidal alignment picture and may be related to recent observations showing that satellite galaxies are preferentially located along the axis of surrounding filaments \cite{tempel15}. Conversely, satellite alignment appears to be significantly weaker \cite{sifon14}, although the most luminous satellites may be somewhat aligned with central positions and the surrounding large-scale structure \cite{singh14,zhang13}.

A detailed description of these intra-halo alignments is beyond the scope of this paper. Instead, motivated by the observational results of \cite{singh14} (see figure~\ref{fig:avg_IA1}), we will make the following simple ansatz: within the halo $\avgplust$ saturates at a constant value. The radius at which this saturation occurs will in general depend on the properties of the shape and density tracers under consideration. It can be related to the halo virial radius or be left as a free parameter. In this work, we set $\avgplust$ to a constant value at $R_{200c}$, defined by mean halo density $\bar{\rho}(r<R_{200c}) = 200 \rho_{\rm crit}$. This value is somewhat inside the halo virial radius and corresponds to $R_{200c} = 0.43$\hMpc~for the average LOWZ galaxy halo. Note that the effect of projected correlations from pairs of galaxies not in the same halo rapidly drops for $r_p<R_{\rm halo}$, and thus the pairs of objects that dominate the measured correlations on these scales are indeed in the same halo.

If $\avgplust$ is indeed dominated by central shape alignment with satellite positions, saturating at a constant value implies that the average satellite distribution maintains a reasonably constant shape and orientation as a function of radius. The apparent lack of scale dependence within the halo may be in tension with the behavior of dark matter in simulations, which exhibits significant internal ``twisting'' (e.g.\ \cite{schneider11}). One possibility is that the satellite distribution may maintain a more regular shape than the dark matter. Alternatively, the behavior of $\avgplust$ seen in the LOWZ galaxy sample may not be typical of less massive satellites. Potential systematic uncertainties due to fiber collisions and shape measurement could also affect $\avgplust$ measured on these small scales (see \cite{singh14}), in which case the observed saturation may not be physical. In any case, some degree of saturation or other departure from tidal alignment is necessary to avoid divergent alignment on small scales.

The small-scale alignment behavior of LOWZ galaxies may not be universally valid when considering different halo masses and satellite properties. Further understanding of shape alignments within halos remains an important goal. We suggest that the $\avgplust$ statistic, as well as similar statistics such as those used in \cite{tenneti14b,singh14}, be used to probe alignment around centrals and satellites with different masses and other properties. For instance, the results presented in figures 11-12 of \cite{tenneti14b} show that average alignments of galaxy shapes resolved in hydrodynamical simulations reach a similar saturation within the one halo regime. Further observations ands simulations, including centrals and satellites spanning a wide range in color and luminosity, will be helpful in probing these questions.

\subsection{Contribution to cosmic shear}

IA contributes a potentially important systematic signal to the measured cosmic shear. In previous sections, we have considered IA correlations between biased tracers. However, the relevant intrinsic alignment contamination to cosmic shear measurements is sourced by correlations between the shapes of tracers and the dark matter density field itself. In this section, we will briefly describe tidal alignment modeling of these terms. For a more detailed study of the impact of IA contamination for various lensing surveys, see \cite{krause15inprep}.

The two-point shape correlations can be expressed in terms of the $E$ and $B$ modes of the angular power spectrum (or angular correlation function):
\begin{align}
C^{EE} &= C^{EE,GG}+C^{EE,GI}+C^{EE,II},\notag\\
C^{BB} &= C^{BB,II},
\end{align}
where the $GG$ term is the standard cosmic shear signal, $GI$ is the cross-correlation between lensing shear and IA (which remains even when cross-correlating shapes from different redshift bins), and $II$ is the IA auto-correlation from pairs of galaxies that are affected by the same fluctuations in the gravitational potential. The $GI$ term corresponds to $\langle \delta | \tilde{\gamma}_+\rangle$, while the $II$ term comes from $\langle \tilde{\gamma}_{(+,\times)} | \tilde{\gamma}_{(+,\times)}\rangle$. See \cite{hirata04} for further discussion.

In our perturbative (SPT) treatment of tidal alignment, these correlations are essentially the same as those presented above (section~\ref{sec:combining}), with the galaxy density field $\delta_g$ on the left side of the $GI$ correlations replaced with the dark matter density $\delta$ (the $\delta_g$ appearing in the tracer density weighting remains). Comparing to eq.~\ref{eq:PgE1}, we find at one loop:
\begin{align}
P_{\delta E}(\mathbf{k})&= A(z) \sin^2(\theta_k)\left[P_{0|S}+b_1P_{0|0E} + \mathcal{O}(P_L^3)\right].
\end{align}
The $II$ correlations of eqs.~\ref{eq:PEE1}-\ref{eq:PBB1} are unchanged. See appendix~\ref{sec:app_SPTcalc} for the derivation of these terms.

For our non-perturbative treatment, explicitly separating the effects of alignment and clustering, we must now consider the average alignment of shape tracers sourced by the dark matter density field itself. While the relevant IA auto-correlation ($II$ term) is unchanged from equation~\ref{eq:sep1}, the IA-density cross-correlation ($GI$ term) in configuration space is:
\begin{align}
w_{\delta +}&=\avgplus^{\rm DM}_{r_p}(2\Pi_{\rm max} + w_{g \delta}),\notag\\
&=\frac{1}{2\Pi_{\rm max}}A(z)\left(w_{\delta \delta, 2}+\frac{58}{105}b_1\sigma_S^2 w^{\rm lin}_{\delta \delta, 2}\right)\left(2\Pi_{\rm max} + w_{g \delta}\right).
\end{align}
The quantity $\avgplus^{\rm DM}_{r_p}$ is the average IA correlation sourced by dark matter density, measured at the positions of shape tracers, which leads to the same large-scale enhancement found in eq.~\ref{eq:plust}. Although not directly observable in galaxy shape measurements, this $w_{\delta +}$ correlation is closely related to the GI contribution to cosmic shear and can be directly measured in simulations (e.g.\ \cite{tenneti14b}).

\begin{figure}[h!]
\begin{center}
\resizebox{5.5in}{!}{
\includegraphics{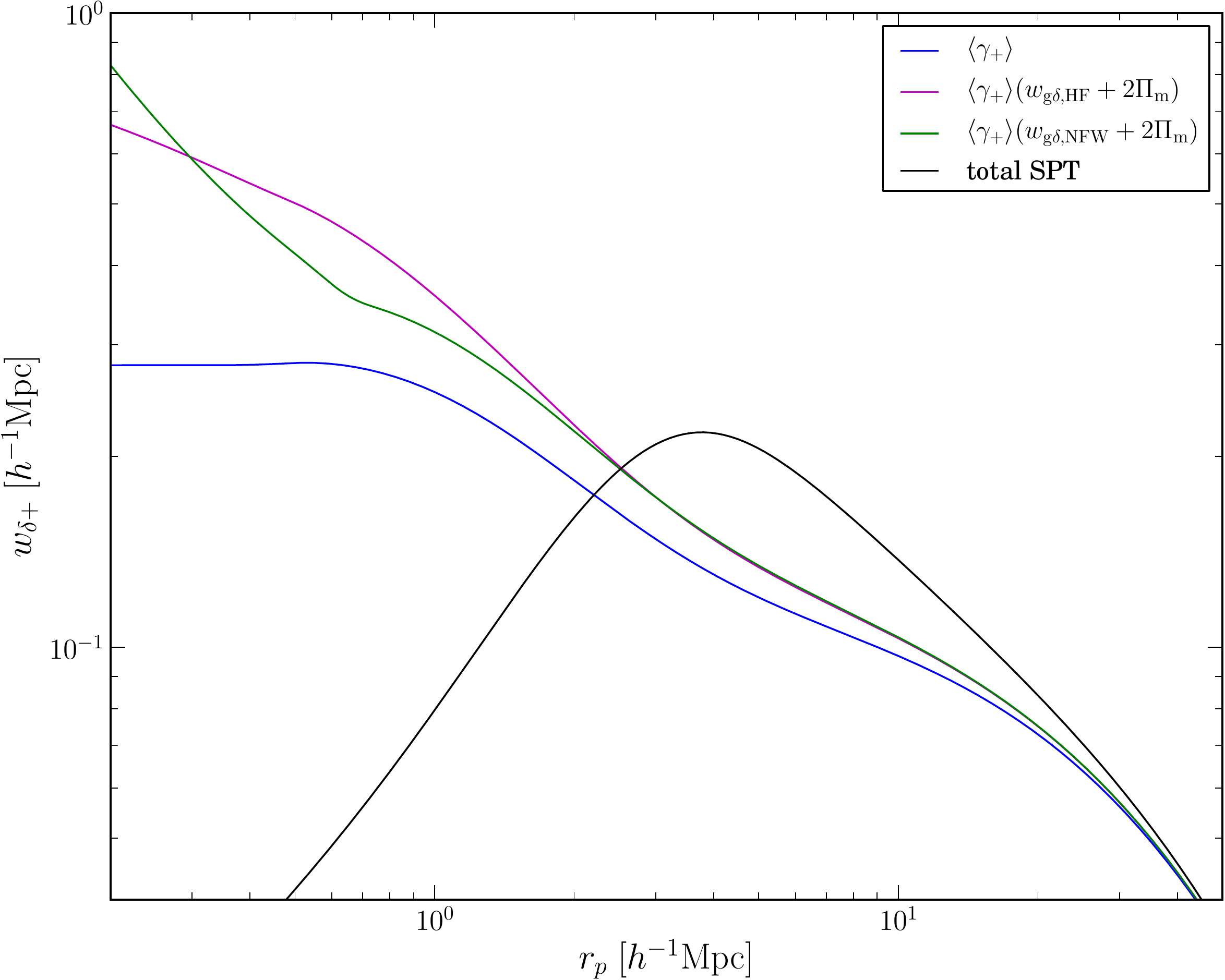}
}
\end{center}
\caption{Predictions for $w_{\delta +}$ are shown for both the SPT and non-perturbative models. Shape tracers are assumed to have the same bias and alignment properties as the LOWZ galaxies \cite{singh14}. As discussed in the text, $\langle \gamma_+\rangle$ saturates below $r_p=R_{200c}$, and results are shown for $w_{g \delta}$ predicted using both Halofit (with linear bias) and an NFW profile. For reference, the $\langle \gamma_+\rangle$ line (blue) shows the alignment effect without shape tracer clustering.}
\label{fig:deltaplus}
\end{figure}

Figure~\ref{fig:deltaplus} shows the tidal alignment predictions for $w_{\delta +}$ (the $GI$ contribution) from shape tracers similar to the LOWZ sample for both the SPT model and the non-perturbative (NFW halo model) approach. For the SPT modeling, we continue to smooth the tidal field at $k_{\rm sm}=1$~\hMpcinv. In the non-perturbative model, the tidal field is not explicitly smoothed, although following the treatment in section~\ref{sec:one_halo}, the value of $\avgplust$ is set to a constant within $R_{\rm const} = R_{200c}$. The prediction for $w_{g \delta}$ is made using both Halofit (with linear bias) and an NFW profile. For illustrative purposes, we consider only dark matter halos at the LOWZ mass. This approach is overly simplified - further investigation on alignments within the one-halo regime as well as including contributions from the entire halo mass range are required.

\section{Discussion}
\label{sec:discussion}

In this work, we have developed the tidal alignment framework for galaxy intrinsic alignments. We included nonlinear contributions from source density weighting, dark matter clustering, and biasing. We have also explored the effect of smoothing the tidal field on the IA correlations. Accounting for these effects improves the consistency of these models beyond the linear and nonlinear alignment (NLA) models and provides a more accurate description of the observed alignments of luminous red galaxies. In their fiducial forms, these models are constructed without introducing additional free parameters (beyond the IA response value $C_1$ and the galaxy bias parameters or relevant halo mass). The other required elements - the smoothing scale and resulting value of $\sigma_S^2$ and, in the case of the one-halo term, the radius where $\avgplus$ becomes constant - can be predicted from physical arguments or left as free parameters within a narrowly allowed range. Since marginalizing over an increased number of free parameters in the IA model will reduce the constraining power of lensing measurements, minimizing additional free parameters is an important advantage to our models.

Of the three sources of nonlinearity that contribute in the tidal alignment picture, source density weighting is the most significant, particularly for highly biased tracers. Density weighting effects arise because we observe intrinsic alignments only at the locations where galaxies form. In the case of tidal alignment, we are thus sampling the tidal field in these special locations. The clustering of shape tracers increases IA correlations on small scales and generates a $B$-mode autocorrelation signal. Somewhat less intuitively, the impact of density weighting extends to large scales, where it effectively enhances the amplitude of the linear theory prediction. This term contributes to an effective value of $C_1$, which depends on both the bias of the shape tracer and the relevant smoothing scale of the tidal field. Physically, this effect is a direct result of sampling the tidal field in the special locations where galaxies form. More highly biased objects tend to form in regions with larger tidal field fluctuations, leading to larger intrinsic ellipticity correlations, even if the underlying response to a given tidal field is the same for all halos/galaxies.

We have also shown that observed shape correlations can be modeled non-perturbatively by separately considering lensing and alignment components. In particular, we developed a straightforward description of alignment on small scales using the tidal field from an NFW halo profile and saturating the value of $\avgplust$ inside the halo. This model is likely overly simplistic - further work in modeling, observations, and simulations are required to better understand these effects. Nevertheless, this model is simple to implement and likely to be more accurate than existing models on smaller scales where a significant amount of lensing information is contained.

The two approaches described in this work are complementary and may be useful in different regimes. The SPT model is straightforward to implement and provides a consistent and well-understood description on intermediate and large scales. Splitting the correlations into alignment and clustering components provides more flexibility in including non-perturbative physics, which allows a description on small scales, including within the one-halo regime. The model's accuracy is then determined by the particular choices for the nonlinear tidal field and galaxy/matter clustering. Both of these approaches can be generalized to include other alignment mechanisms (such as tidal torquing).

As plans are made for analyzing current and future weak lensing experiments, it is important to emphasize that the simplest IA models may not correctly describe all galaxies. Assuming tidal alignment when the true astrophysics of galaxy formation produces more complicated behavior will yield systematic bias in the final parameter constraints. Moreover, even forecasting the impact of IA requires a model both sufficiently accurate and complex to avoid overestimating the statistical power of a given survey and mitigation scheme \cite{krause15inprep}. In this paper, we have neglected contributions from higher order dependence on the tidal field (i.e.\ tidal torquing models). Although symmetry allows such terms, and they are likely present at some level for galaxies kinematically dominated by angular momentum, current observations provide no strong evidence that these terms are significant. Nevertheless, until higher-order alignment contributions can be quantified or more tightly constrained in observation or realistic simulation, it will be important to include them. In future work, we plan to integrate these terms into a consistent framework.

The tidal alignment picture has been successful in describing IA in both astrophysical observations and simulations. In consistently including nonlinear contributions, this work should improve the accuracy and flexibility of tidal alignment as a tool in precision cosmology. Furthermore, studying these contributions probes astrophysical effects which must be understood as we develop a more complete model of intrinsic galaxy shapes.
\acknowledgments

We thank Chris Hirata, Rachel Mandelbaum, David Weinberg, and Benjamin Joachimi for helpful discussions. We also thank Sukhdeep Singh and Teppei Okumura for sharing and discussing their measurements. J.B.\ acknowledges the support of a CCAPP fellowship. U.S.\ is supported in part by the NASA ATP grant NNX12AG71G. Z.V.\ is supported in part by the U.S.\ Department of Energy contract to SLAC no.\ DE-AC02-76SF00515.

\clearpage
\appendix 
\section{Calculating nonlinear contributions in SPT}
\label{sec:app_SPTcalc}
Details of the calculations for $P_{gE}$, $P_{EE}$, $P_{BB}$ are presented below.
\subsection{Calculating $P_{gE}$}
We start by determining all possible correlations which can contribute to $P_{gE}$:
\begin{align}
\label{eq:P_gE_all}
\langle \delta_g | \tilde{\gamma}_{+}\rangle = &
 A(z) \times\Big[
 (1-b_2\sigma^2) b_1 f_+\langle \delta | \delta\rangle + b_1^2 \langle \delta | \delta \circ (f_+\delta) \rangle + b_2 f_+ \langle \delta \circ \delta | \delta \rangle \notag\\
 &+ b_1 b_2 \langle \delta | (\delta\circ \delta) \circ (f_+ \delta) \rangle + b_3 f_+ \langle (\delta \circ \delta \circ \delta) | \delta \rangle
 + b_1 b_2  \langle \delta \circ \delta | \delta \circ (f_+ \delta) \rangle
 +\ldots\Big],
\end{align}
where dots represent the terms that contribute at higher than one-loop order. We remove terms that are zero by symmetry, and by decomposing the remaining correlators we find:
\begin{align}
\label{eq:P_gE_decomp}
\langle \delta_g | \tilde{\gamma}_{+}\rangle = & A(z) \times\Big[
  \big( (1-b_2\sigma^2) b_1 +b_1b_2 \sigma^2 +3 b_3 \sigma^2 ) f_+\langle \delta | \delta \rangle 
  + b_1^2 \langle \delta | \delta \circ (f_+\delta)\rangle + b_2  f_+ \langle \delta \circ \delta | \delta \rangle \notag\\ 
  &+ b_1 b_2 \langle \delta | (\delta\circ \delta) \circ (f_+ \delta) \rangle_c + b_3 f_+ 
  \langle (\delta \circ \delta \circ \delta) | \delta \rangle_c + b_1 b_2 \langle \delta \circ \delta | \delta \circ (f_+ \delta) \rangle_c
+\ldots\Big].
\end{align}
The ``$c$'' subscript denotes the tightly coupled component of a correlator (i.e.\ components not proportional to $\sigma^2$). After the bias renormalization $b_1\rightarrow b_1+3b_3\sigma^2$ (e.g.\ \citep{mcdonald06, assassi14}), and keeping all remaining one-loop terms (i.e.\ $\mathcal{O}(\delta_{\rm lin}^4)$) we find: 
\begin{align}
\label{eq:leading}
\langle \delta_g | \tilde{\gamma}_{+}\rangle =  A(z) \times\Big[
  &b_1 f_+\langle \delta | \delta \rangle + b_1^2 \langle \delta | \delta \circ (f_+\delta)\rangle +b_2 f_+ \langle \delta \circ \delta | \delta \rangle\notag\\
 &+b_1b_2 \langle \delta \circ \delta | \delta \circ (f_+ \delta) \rangle_c
 + \mathcal{O}(\delta_{\rm lin}^6)\Big].
\end{align}
Although derived through perturbative considerations, this model for $P_{gE}$ can now be considered more generally, and each term can be evaluated to arbitrary precision. This is possible since each term has a unique bias dependence, and the corresponding correlators are 
IR safe \cite{carrasco14}. While the particular bias renormalization scheme will change depending on the choice of PT technique (or loop order), the overall form will not.

We now consider the correlators from equation \ref{eq:leading}. Note that the resulting convolution integrals often include the IA operators $f_+$ and $f_{\times}$. In appendix~\ref{sec:app_angular_int}, we derive the relevant angular integrations over these operators.

\subsubsection{The $\langle \delta | \delta \rangle$ term}
\label{sec:P00}
Expanding this term at one-loop order:
\begin{align}
\langle \delta | \delta \rangle = \langle \delta^{(1)} | \delta^{(1)} \rangle+\langle \delta^{(2)} | \delta^{(2)} \rangle +2\langle \delta^{(1)} | \delta^{(3)} \rangle.
\end{align}
In terms of the power spectrum with the standard definition, $(2\pi)^3 P(k)_{0|0}\delta^D(\mathbf{k}+\mathbf{k}')=\langle \delta | \delta \rangle$, we have:
\begin{align}
P_{0|0}(k) &= P_L(k)+2 I_{00}(k)+6 k^2 J_{00}(k)P_L(k) ,
\end{align}
where $I_{00}$ and $J_{00}$ are the one-loop SPT corrections to the matter power spectrum (e.g.\ \cite{vlah12}):
\begin{align}
I_{00}(k)&=\int \frac{d^3q}{(2\pi)^3}P_L(q)P_L(|\mathbf{k}-\mathbf{q}|)F^2_2(\mathbf{q},\mathbf{k-q}),\\
J_{00}(k)&=\frac{1}{k^2}\int \frac{d^3q}{(2\pi)^3}P_L(q)F_3(\mathbf{k},\mathbf{q},\mathbf{-q}).
\notag
\end{align}
We have employed the relevant gravity kernels:
\begin{align}
F_2(\mathbf{q},\mathbf{k-q})&= \frac{7\mu + 3 \alpha -10\alpha\mu^2}{14\alpha(1+\alpha^2-2\alpha\mu)},\\
F_3(\mathbf{k},\mathbf{q},\mathbf{-q})&= \frac{1}{3024q^2}\left[ \frac{12}{\alpha^2}-158+100\alpha^2-42\alpha^4
+\frac{3}{\alpha^3}(\alpha^2-1)^3(7\alpha^2+2)\ln\left(\frac{\alpha+1}{|\alpha-1|}\right)\right],\notag
\end{align}
where $\alpha=q/k$ and $\mu=\hat{k}\cdot\hat{q}$. The smoothing of the density field easily factors out, yielding: $P_{0|S}(k) = S(k) P_{0|0}(k)$ and $P_{S|S}(k) = S^2(k)P_{0|0}(k)$. Although this expression for $P_{0|0}(k)$ is correct at one-loop order, it is known to provide a poor description on comparatively large scales. In this work, we instead use the Halofit fitting formula \cite{smith03} for nonlinear $P_{0|0}(k)$.

\subsubsection{The $\langle \delta | \delta \circ (f_i\delta)\rangle$ term}

Expanding the term to one-loop order:
\begin{align}
\langle \delta | \delta \circ (f_+\delta)\rangle = \langle \delta^{(2)} | \delta^{(1)} \circ (f_+\delta^{(1)}) \rangle
+\langle \delta^{(1)} | \delta^{(2)} \circ (f_+\delta^{(1)}) \rangle+\langle \delta^{(1)} | \delta^{(1)} \circ (f_+\delta^{(2)}) \rangle.
\end{align}
These correlators correspond to contributions to the power spectrum:
\begin{align}
P_{0|0E}(k)= A_{0|0E}(k)+B_{0|0E}(k)+C_{0|0E}(k).
\end{align}
These terms can be written:
\begin{align}
  A_{0|0E}(k)&=2\int \frac{d^3q}{(2\pi)^3} f_+(-\mathbf{q}) F_2(\mathbf{k}-\mathbf{q},\mathbf{q}) P_L(q)P_L(|\mathbf{k}-\mathbf{q}|),\notag\\
  B_{0|0E}(k)&=2P_L(k)\int \frac{d^3q}{(2\pi)^3} f_+(-\mathbf{q}) F_2(-\mathbf{k},\mathbf{q}) P_L(q),\notag\\
  C_{0|0E}(k)&=2P_L(k)\int \frac{d^3q}{(2\pi)^3} f_+(-\mathbf{k}+\mathbf{q}) F_2(-\mathbf{k},\mathbf{q}) P_L(q) .
\label{eq:ddfd_ABC_v1}
\end{align}
Using eq.~\ref{eq:finqbasis} for the integral over $f_+$ and preforming the coordinate transformation $\mathbf{q}\rightarrow \mathbf{k} -\mathbf{q}$ in $C_{0|0E}(k)$ yields:
\begin{align}
  A_{0|0E}(k)&=2\sin^2(\theta_k)\int \frac{d^3q}{(2\pi)^3} \mathcal{P}_2(\mu_q) S(q) F_2(\mathbf{k}-\mathbf{q},\mathbf{q}) P_L(q) P_L(|\mathbf{k}-\mathbf{q}|) , \notag\\
  B_{0|0E}(k)&=\sin^2(\theta_k)\frac{8}{105}\sigma_S^2 P_L(k) , \notag\\
  C_{0|0E}(k)&=2\sin^2(\theta_k) P_L(k) \int \frac{d^3q}{(2\pi)^3} \mathcal{P}_2(\mu_q) S(q) F_2(-\mathbf{k},\mathbf{k}-\mathbf{q})P_L(|\mathbf{k}-\mathbf{q}|) ,
\label{eq:ddfd_ABC_v2}
\end{align}
where $\sigma^2=\int\frac{d^3q}{(2\pi)^3} P_L(q)$ and $\sigma_S^2=\int\frac{d^3q}{(2\pi)^3} P_L(q) S(q)$. Combining these terms, we have:
\begin{align}
\label{eq:P001}
P(k)_{0|0E}=& \frac{58}{105}\sigma_S^2 P_L(k)+ 2\int \frac{d^3q}{(2\pi)^3} \mathcal{P}_2(\mu_q) S(q) F_2(\mathbf{k}-\mathbf{q},\mathbf{q}) P_L(q)P_L(|\mathbf{k}-\mathbf{q}|)\notag\\
&+2 P_L(k) \int \frac{d^3q}{(2\pi)^3} S(q) \left[ \mathcal{P}_2(\mu_q) F_2(-\mathbf{k},\mathbf{k}-\mathbf{q})P_L(|\mathbf{k}-\mathbf{q}|) -\frac{5}{21} P_L(q)\right].
\end{align}
In the last step we have absorbed the $k\rightarrow0$ contribution of $C_{0|0E}$ into the $\sigma_S^2 P_L(k)$ term. Note that evaluation of the $C_{0|0E}(k)$ term in \ref{eq:ddfd_ABC_v2} is more straightforward in the original coordinates of eq.~\ref{eq:ddfd_ABC_v1}.

\subsubsection{The $\langle \delta \circ \delta | \delta \rangle$ term}

Expanding the term to one-loop order:
\begin{align}
\langle \delta \circ \delta | \delta \rangle = \langle \delta^{(1)} \circ \delta^{(1)} | \delta^{(2)} \rangle +2 \langle \delta^{(2)} \circ \delta^{(1)} | \delta^{(1)} \rangle.
\end{align}
As with $P_{0|S}(k)$, the smoothing filter factors out:
\begin{align}
P_{00|S}(k)= S(k) P_{00|0}(k) = S(k)\left[A_{00|0}(k)+2B_{00|0}(k)\right],
\end{align}
where we have:
\begin{align}
  A_{00|0}(k)&=2\int \frac{d^3q}{(2\pi)^3} F_2(\mathbf{k}-\mathbf{q},\mathbf{q}) P_L(q)P_L(|\mathbf{k}-\mathbf{q}|) ,\notag\\
  B_{00|0}(k)&=2P_L(k) \int \frac{d^3q}{(2\pi)^3} F_2(-\mathbf{k},-\mathbf{q}) P_L(q)=\frac{34}{21}\sigma^2 P_L(k) .
\end{align}
Bringing everything together, we find:
\begin{align}
P_{00|S}(k)= S(k) \left[\frac{68}{21}\sigma^2 P_L(k) + 2\int \frac{d^3q}{(2\pi)^3} F_2(\mathbf{k}-\mathbf{q},\mathbf{q}) P_L(q)P_L(|\mathbf{k}-\mathbf{q}|)\right].
\end{align}

\subsubsection{The $\langle  \delta \circ \delta | \delta \circ (f_i\delta)\rangle_c$ term}

Expanding the term to one-loop order:
\begin{align}
\langle \delta \circ \delta | \delta \circ (f_+\delta)\rangle = \langle \delta^{(1)} \circ \delta^{(1)} | \delta^{(1)} \circ (f_+\delta^{(1)}) \rangle.
\end{align}
This leads to the power spectrum:
\begin{align}
P_{00|0E}(k)= 2\int \frac{d^3q}{(2\pi)^3} f_+(-\mathbf{q}) P_L(q)P_L(|\mathbf{k}-\mathbf{q}|) .
\end{align}
Using eq.~\ref{eq:finqbasis} yields:
\begin{align}
P_{00|0E}(k)= 2\sin^2(\theta_k) \int \frac{d^3q}{(2\pi)^3} \mathcal{P}_2(\mu_q)  S(q) P_L(q)\big( P_L(|\mathbf{k}-\mathbf{q}|) \big).
\end{align}
Note that the $k\rightarrow 0$ limit is zero due to the quadrupolar dependence on $\mu_q$.

\subsubsection{Combining terms}

Putting all the terms from eq.~\ref{eq:leading} together yields:
\begin{align}
P_{gE}(k,\theta_k) = & A(z) \sin^2(\theta_k)\times\Big[
  b_1 P_{0|S}(k)
+b_1^2 \frac{58}{105}\sigma_S^2 P_L(k)\notag\\
&+ 2 b_1^2 \int \frac{d^3q}{(2\pi)^3} \mathcal{P}_2(\mu_q) F_2(\mathbf{k}-\mathbf{q},\mathbf{q}) S(q) P_L(q)P_L(|\mathbf{k}-\mathbf{q}|)\notag\\
&+2 b_1^2 P_L(k) \int \frac{d^3q}{(2\pi)^3} S(q) \left[\mathcal{P}_2(\mu_q) F_2(-\mathbf{k},\mathbf{k}-\mathbf{q})P_L(|\mathbf{k}-\mathbf{q}|) -
\frac{5}{21} P_L(q)\right]\notag\\
&+ 2 b_2 \int \frac{d^3q}{(2\pi)^3}  F_2(\mathbf{k}-\mathbf{q},\mathbf{q}) S(q) P_L(q)P_L(|\mathbf{k}-\mathbf{q}|)\notag\\
&+ 2 b_1b_2 \int \frac{d^3q}{(2\pi)^3} \mathcal{P}_2(\mu_q)  S(q) P_L(q)\big( P_L(|\mathbf{k}-\mathbf{q}|) \big)
+\mathcal{O}(P_{L}^3)\Big],
\end{align}
where we have absorbed the divergent $\sigma^2$ contribution into the definition of $b_1$ with the bias renormalization:
\begin{align}
b_1 \rightarrow b_1 + \frac{68}{21}b_2\sigma^2 .
\end{align}
Combined with the earlier bias renormalization ($b_1\rightarrow b_1+3b_3\sigma^2$), we have:
\begin{align}
b_1 \rightarrow b_1 + \left(3 b_3 + \frac{68}{21}b_2\right)\sigma^2 ,
\end{align}
which is consistent with previous results (e.g.\ \citep{mcdonald06}).

\subsection{Calculating $P_{EE}$ and $P_{BB}$}
Using our freedom to choose $k_y=0$, the $E$- and $B$-mode autocorrelation power spectra are given by:
\begin{align}
\langle \tilde{\gamma}_+ | \tilde{\gamma}_+ \rangle=&A^2(z)\times \Big[(1-2b_2\sigma^2)f^2_+\langle \delta | \delta \rangle
+2b_1 f_+ \langle \delta | \delta \circ (f_+\delta)\rangle \notag\\
&+2b_2 f_+ \langle \delta | \delta \circ \delta \circ (f_+\delta)\rangle
+b_1^2\langle \delta \circ (f_+\delta)| \delta \circ (f_+\delta)\rangle
+\mathcal{O}(\delta_{\rm lin}^6)\Big]\notag,\\
\langle \tilde{\gamma}_\times | \tilde{\gamma}_\times \rangle=&A^2(z)\times \Big[b_1^2\langle \delta \circ (f_\times\delta)| \delta \circ (f_\times\delta)\rangle
+\mathcal{O}(\delta_{\rm lin}^6)\Big],
\end{align}
where we have used $f_{\times}(k_y=0)=0$. Separating into disconnected and tightly coupled correlators, we find:
\begin{align}
\langle \tilde{\gamma}_+ | \tilde{\gamma}_+ \rangle=&A^2(z)\times \Big[f^2_+\langle \delta | \delta \rangle
+2b_1 f_+ \langle \delta | \delta \circ (f_+\delta)\rangle
+b_1^2\langle \delta \circ (f_+\delta)| \delta \circ (f_+\delta)\rangle
+\mathcal{O}(\delta_{\rm lin}^6)\Big]\notag,\\
\langle \tilde{\gamma}_{\times} | \tilde{\gamma}_{\times} \rangle=&A^2(z)\times \Big[b_1^2\langle \delta \circ (f_\times\delta)| \delta \circ (f_\times\delta)\rangle\notag
+\mathcal{O}(\delta_{\rm lin}^6)\Big].
\end{align}
The $\langle \delta | \delta \rangle$ and $\langle \delta | \delta \circ (f_+\delta)\rangle$ terms have been described above. As expected, the leading $B$-mode contribution appears at one-loop and is due to weighting by the shape tracer density.
\subsubsection{The $\langle \delta \circ (f_i\delta) | \delta \circ (f_i\delta)\rangle$ term}
At one-loop, the only contribution to this term is $\langle \delta^{(1)} \circ (f_i\delta^{(1)}) | \delta^{(1)} \circ (f_i\delta^{(1)})\rangle$. The $E$-mode correlation is:
\begin{align}
P_{0E|0E}(\mathbf{k})= \int \frac{d^3q}{(2\pi)^3}  \left[f_E(\hat{q}) f_E(\hat{q_2})+ f_E(-\hat{q})f_E(\hat{q})\right] P_L(q) P_L(q_2),
\end{align}
where $\mathbf{q_2}=\mathbf{k}-\mathbf{q}$. $P_{0B|0B}(\mathbf{k})$ is given by the equivalent expression with $f_E \rightarrow f_B$. Using eq.~\ref{eq:finqbasis_limber}, and making the Limber approximation ($\theta_k=\pi/2)$:
\begin{align}
P_{0E|0E}(k)&= \int \frac{d^3q}{(2\pi)^3}  \left[\left(\frac{3 - 14 \mu_q^2 + 19 \mu_q^4}{8}\right)P_L(q)S^2(q) \left[P_L(q_2)-2P_L(q)\right]\right.\notag\\
&\left.+\left(\frac{8 \alpha \mu_q (3 \mu_q^2-1) + 4 (-1 + 3 \mu_q^2) + \alpha^2 (3 - 14 \mu_q^2 + 19 \mu_q^4)}{8 (1 + \alpha^2 - 2 \alpha \mu_q)}\right)
P_L(q) P_L(q_2)S(q)S(q_2)\right],\notag\\
P_{0B|0B}(k)&= \int \frac{d^3q}{(2\pi)^3}  \Biggl[\left(  2 \mu_q^2 (1 - \mu_q^2) \right)P_L(q)S^2(q)\left(P_L(q_2)-2P_L(q)\Biggr)\right.\notag\\
&\Biggl.+  \left(\frac{2 \alpha \mu_q(\mu_q-1) (1 - \alpha \mu_q)}{k^2 + q^2 - 2 k q \mu_q} \right) P_L(q) P_L(q_2)S(q)S(q_2)\Biggr],
\end{align}
where $\alpha=q/k$. Smoothing affects these terms on large-scales because of a zero-lag (scale-independent) contribution, which comes from a convolution over all scales (including the smoothed small scales).
\subsubsection{Combining terms}
Combining all contributions to the IA autocorrelation at one loop, we have:
\begin{align}
P_{EE}(\mathbf{k})&=A^2(k)\left[\sin^4(\theta_k)S^2(k)P_{0|0}(k)+2 b_1\sin^4(\theta_k)S(k)P_{0|0E}(k)\right.\notag\\
&~~~~~~~~~~~~~~~~\left.+ b_1^2 P_{0E|0E}(k,\theta_k)+\mathcal{O}(P_L^3)\right],\notag\\
P_{BB}(\mathbf{k})&=A^2(k)\left[b_1^2 P_{0B|0B}(k,\theta_k)+\mathcal{O}(P_L^3)\right].
\end{align}
\section{Simplifying the angular integration of IA operators}
\label{sec:app_angular_int}
As seen in appendix~\ref{sec:app_SPTcalc}, the SPT expansion leads to convolutions of the linear power spectrum, similar in form to:
\begin{align}
P(\mathbf{k}) \sim \frac{1}{(2\pi)^3}\int d^3q\, f_i(\mathbf{q}) F(\mathbf{k-q},\mathbf{q})P(q)P(|\mathbf{k-q}|),
\end{align}
where $F$ denotes a gravity kernel, and $f_i$ is the relevant IA operator. We have the freedom to choose an arbitrary coordinate system for the wavevector $\mathbf{q}$. If we rotate to a coordinate system in which $\hat{k}$ is the polar axis, the only azimuthal dependence in the integrand is in $f_i(\mathbf{q})$. We can then analytically evaluate the $\phi_q$ part of the integral. In this appendix we show the results of the azimuthal integrals of the relevant IA operators. We define $\mathbf{q_2}\equiv\mathbf{k}-\mathbf{q}$ and use the freedom to set $k_y=0$ in the final results.

For convolution integrals with only one IA operator, only $f_+$ is relevant.
\begin{align}
\label{eq:finqbasis}
\frac{1}{2\pi}\int\limits_0^{2\pi}d\phi_q f_+(\mathbf{q})&=\frac{1}{2}(3\mu_q^2-1)f(\hat{k}) = \Leg_2(\mu_q)\sin^2(\theta_k),\notag\\
\frac{1}{2\pi}\int\limits_0^{2\pi}d\phi_q f_+(\mathbf{q_2})&=\frac{1+\alpha^2 \Leg_2(\mu_q) - 2\alpha\mu_q }{1+\alpha^2 -2\alpha\mu_q } \sin^2(\theta_k),
\end{align}
where $\Leg_i$ are the Legendre polynomials, $\alpha=q/k$, $\mu_q \equiv \hat{q}\cdot\hat{k}=\cos\theta_q$.

For convolution integrals with two IA operators, the $\theta_k$-dependence does not factor out of resulting expressions. For simplicity, we make the Limber approximation, in which only transverse modes ($\theta_k=\pi/2$) contribute:
\begin{align}
\label{eq:finqbasis_limber}
\frac{1}{2\pi}\int\limits_0^{2\pi}d\phi_q f_+(\mathbf{q}) f_+(\mathbf{q})  &= \frac{1}{8} (3 - 14 \mu_q^2 + 19 \mu_q^4),\notag\\
\frac{1}{2\pi}\int\limits_0^{2\pi}d\phi_q f_+(\mathbf{q}) f_+(\mathbf{q_2})  &= \frac{8 \alpha \mu_q (1-3 \mu_q^2) + 4 (3 \mu_q^2-1) + \alpha^2 (3 - 14 \mu_q^2 + 19 \mu_q^4)}{8 (1 + \alpha^2 - 2 \alpha \mu_q)},\notag\\
\frac{1}{2\pi}\int\limits_0^{2\pi}d\phi_q f_+(\mathbf{q_2}) f_+(\mathbf{q_2})  &= \frac{1 - 4 \alpha \mu_q + 2 \alpha^3 \mu_q (1 - 3 \mu_q^2)}{(1 + \alpha^2 - 2 \alpha \mu_q)^2} \notag\\
&+ \frac{8 \alpha^2 (-1 + 7 \mu_q^2) + \alpha^4 (3 - 14 \mu_q^2 + 19 \mu_q^4)}{8 (1 + \alpha^2 - 2 \alpha \mu_q)^2},\notag\\
\frac{1}{2\pi}\int\limits_0^{2\pi}d\phi_q f_{\times}(\mathbf{q}) f_{\times}(\mathbf{q})  &= 2 \mu_q^2 (1 - \mu_q^2),\notag\\
\frac{1}{2\pi}\int\limits_0^{2\pi}d\phi_q f_{\times}(\mathbf{q}) f_{\times}(\mathbf{q_2})  &= \frac{2 \alpha \mu_q(\mu^2_q-1) (1 - \alpha \mu_q)}{1 + \alpha^2 - 
 2 \alpha \mu_q},\notag\\
\frac{1}{2\pi}\int\limits_0^{2\pi}d\phi_q f_{\times}(\mathbf{q_2}) f_{\times}(\mathbf{q_2})  &= \frac{2 \alpha^2 (1-\mu^2_q) (1 - \alpha \mu_q)^2}{(1 + \alpha^2 -   2 \alpha \mu_q)^2}.
\end{align}
Note that because the IA operators are even, in the $k \rightarrow 0$ limit, these expressions are unchanged under the exchange $\mathbf{q} \leftrightarrow \mathbf{q_2}$.

\providecommand{\href}[2]{#2}\begingroup\raggedright\endgroup

\end{document}